\documentclass[lettersize,journal]{IEEEtran}
\usepackage{amsmath,amssymb,amsfonts}
\usepackage{algorithmic}
\usepackage{array}
\usepackage[caption=false,font=normalsize,labelfont=sf,textfont=sf]{subfig}
\usepackage{textcomp}
\usepackage{stfloats}
\usepackage{url}
\usepackage{verbatim}
\usepackage{graphicx}
\def\BibTeX{{\rm B\kern-.05em{\sc i\kern-.025em b}\kern-.08em
		T\kern-.1667em\lower.7ex\hbox{E}\kern-.125emX}}
\usepackage{balance}

\usepackage{cite}
\usepackage{xcolor}

\usepackage{subfloat}
\usepackage{epstopdf}

\begin{document}

\include{header}

\title{Over-the-Air Computation in OFDM Systems with Imperfect Channel State Information}

\author{Yilong Chen, Huijun Xing, Jie Xu, Lexi Xu, and Shuguang Cui
	\thanks{Part of this paper has been presented at the IEEE 23rd International Workshop on Signal Processing Advances in Wireless Communication (SPAWC), Oulu, Finland, 4 - 6 July, 2022 \cite{chen2022over}.}
	\thanks{Y. Chen, H. Xing, J. Xu, and S. Cui are with the School of Science and Engineering (SSE) and the Future Network of Intelligence Institute (FNii), The Chinese University of Hong Kong (Shenzhen), Shenzhen, China (e-mail: yilongchen@link.cuhk.edu.cn, huijunxing@link.cuhk.edu.cn, xujie@cuhk.edu.cn, shuguangcui@cuhk.edu.cn).}
	\thanks {L. Xu is with the China Unicom Research Institute, Beijing, China (e-mail: davidlexi@hotmail.com).}
	\thanks{J. Xu is the corresponding author.}
}

\maketitle

\begin{abstract}
	This paper studies the over-the-air computation (AirComp) in an orthogonal frequency division multiplexing (OFDM) system with imperfect channel state information (CSI), in which multiple single-antenna wireless devices (WDs) simultaneously send uncoded signals to a multi-antenna access point (AP) for distributed functional computation over multiple subcarriers. In particular, we consider two scenarios with best-effort and error-constrained computation tasks, with the objectives of minimizing the average computation mean squared error (MSE) and the computation outage probability over the multiple subcarriers, respectively. Towards this end, we jointly optimize the transmit coefficients at the WDs and the receive beamforming vectors at the AP over subcarriers, subject to the maximum transmit power constraints at individual WDs. First, for the special case with a single receive antenna at the AP, we propose the semi-closed-form globally optimal solutions to the two problems using the Lagrange-duality method. It is shown that at each subcarrier, the WDs' optimized power control policy for average MSE minimization follows a regularized channel inversion structure, while that for computation outage probability minimization follows an on-off regularized channel inversion, with the regularization dependent on the transmit power budget and channel estimation error. Next, for the general case with multiple receive antennas at the AP, we present efficient algorithms based on alternating optimization and convex optimization to find converged solutions to both problems. It is shown that with finite receive antennas at the AP, a non-zero computation MSE for AirComp is inevitable due to the channel estimation errors even when the transmit powers at WDs tend to infinity, while with massive receive antennas, the average MSE and outage probability vanish when the channel vectors are independent and identically distributed. Finally, numerical results are provided to demonstrate the effectiveness of the proposed designs.
\end{abstract}


\section{Introduction}

The advancements in artificial intelligence (AI) and the Internet of Things (IoT) are expected to enable numerous new applications, such as smart cities and auto-driving. The emergence of these applications introduces new requirements for wireless data aggregation (WDA), in which distributed data from wireless devices (WDs) need to be aggregated at fusion centers for distributed sensing, distributed edge machine learning (e.g., federated edge learning (FEEL)), and distributed consensus \cite{zhu2021over}. This thus calls for a paradigm shift in the multiple access techniques from the conventional separate-communication-and-computation design to the new integrated-communication-and-computation design. To address this, over-the-air computation (AirComp) has attracted growing research interests for WDA towards beyond fifth-generation (B5G) and six-generation (6G) wireless networks \cite{zhu2021over}, which exploits the waveform superposition properties of wireless channels to facilitate the distributed functional computation among separate WDs. In particular, by implementing AirComp over a multiple access channel (MAC), the access point (AP) receiver is able to directly compute a class of so-called nomographic functions (e.g., arithmetic mean, weighted sum, geometric mean, polynomial, and Euclidean norm) \cite{goldenbaum2014nomographic, abari2016over}, based on distributed data from multiple WD transmitters via their one-shot transmission. For instance, AirComp is particularly appealing for supporting one-shot gradient/model aggregation in FEEL, thereby enabling over-the-air FEEL (Air-FEEL) \cite{zhu2019broadband, zhu2021one, sahin2021distributed}.

In general, AirComp can be realized in both coded and uncoded manners \cite{zhu2021over}. While there have been some prior works studying coded AirComp exploiting structured codes for reliable functional computation (e.g., \cite{nazer2007computation} and \cite{wu2019computation}), uncoded AirComp has recently attracted extensive research interests \cite{cao2020optimized, zhang2021gradient, cao2022optimized, cao2022transmission, zhu2018mimo, zhai2021hybrid, zhong2022over, jiang2020achieving} due to its simplicity in implementation and its optimality in computation mean squared error (MSE) minimization over Gaussian MAC with independent and identically distributed (i.i.d.) Gaussian sources \cite{gastpar2008uncoded}. Despite the benefits, uncoded AirComp also faces various technical challenges. In particular, how to combat against computation errors caused by channel fading as well as noise and interference is particularly challenging. To tackle this issue, there have been various prior works investigating the transmit power control and the transmit/receive beamforming for minimizing the computation errors for AirComp. For instance, the authors in \cite{cao2020optimized} presented the transmit power control design for AirComp over a single-antenna fading MAC, in which the average computation MSE is minimized by properly balancing the tradeoff between signal misalignment and noise-induced errors. Such transmission power control strategy was then extended to accelerate the convergence of model training in Air-FEEL \cite{zhang2021gradient, cao2022optimized, cao2022transmission}. Furthermore, the authors in \cite{zhu2018mimo} utilized the multiple-input multiple-output (MIMO) technique for multimodal AirComp, in which the spatial multiplexing and array gains of MIMO are exploited to simultaneously compute multiple function values with reduced computation MSE. In addition, \cite{zhai2021hybrid} studied the design of hybrid analog and digital beamforming for massive MIMO AirComp, \cite{zhong2022over} proposed the joint transceiver beamforming and device selection design for Air-FEEL with multiple tasks, and \cite{jiang2020achieving} exploited the cooperative diversity for enhancing the outage performance of AirComp via relaying.

Despite the benefits, the implementation of power control and beamforming in AirComp highly depends on the availability of channel state information (CSI) at the WD transmitters and the AP receiver. 
In practice, the CSI can be obtained at the AP and the WDs via proper channel estimation (e.g., by exploiting the uplink-downlink channel reciprocity in time division duplex (TDD) systems). This, however, may introduce channel estimation errors \cite{yoo2006capacity} that would degrade the AirComp performance. In the literature, there have been extensive prior works investigating the impact of imperfect CSI on the design of conventional wireless communications (see, e.g., \cite{eraslan2013performance}), but there are only a handful of prior works analyzing the effect of imperfect CSI on the AirComp performance under different setups \cite{yu2020optimizing, jung2021performance, zhang2022worst, zhai2022beamforming}. For instance, the authors in \cite{zhang2022worst} and \cite{zhai2022beamforming} studied the intelligent reflecting surface (IRS)-assisted AirComp systems with imperfect CSI, in which the joint resource allocation for both transceivers and IRS was designed in \cite{zhang2022worst} for optimizing the worst-case MSE performance and the two-stage stochastic optimization was implemented in \cite{zhai2022beamforming} for performance optimization with reduced signaling overhead. Nevertheless, these prior works \cite{yu2020optimizing, jung2021performance, zhang2022worst, zhai2022beamforming} considered the narrowband transmission with flat-fading channels, and they only provided generally suboptimal solutions under their respective setups. 

Different from prior works, this paper investigates the uncoded AirComp in single-input multiple-output (SIMO) orthogonal frequency division multiplexing (OFDM) systems operating over frequency-selective fading channels, where multiple single-antenna WDs simultaneously transmit uncoded data to a multi-antenna AP for computing function values over subcarriers. Under this setup, we study the joint transmit power control at the WDs and the receive strategy at the AP for optimizing the AirComp performance in terms of minimizing the computation MSE, by taking into account the imperfect CSI caused by channel estimation errors. The computation MSE consists of three components, namely the signal misalignment error, CSI-related error, and noise-induced error. This is in sharp contrast to the case with perfect CSI, in which the CSI-related error is absent, thus making conventional AirComp designs with perfect CSI (e.g., \cite{cao2020optimized, zhu2018mimo, zhai2021hybrid}) inapplicable.
In particular, we consider two different AirComp scenarios with best-effort and error-constrained computation tasks, respectively.
In the first scenario with best-effort computation, the AP aims to compute a given number of function values over a specified time duration (e.g., for gradient/model aggregation in Air-FEEL systems \cite{zhu2019broadband, zhu2021one, sahin2021distributed}), and the batch of function values is treated as a whole (e.g., for the model update in Air-FEEL). In this scenario, a certain level of computation error for a particular function is tolerable. As a result, we dedicate our best efforts to minimizing the average MSE over all subcarriers.
In the second scenario with error-constrained computation, each function value plays a critical role in mission-critical applications (e.g., for distributed consensus in vehicle platooning). In this scenario, we need to ensure the desired error performance for each individual function computation. Consequently, we aim to minimize the computation outage probability or maximize the number of successfully computed functions over all subcarriers, where an outage or unsuccessful computation happens in a subcarrier if the corresponding computation MSE exceeds a specific threshold.

In the aforementioned two scenarios, we minimize the average computation MSE and the computation outage probability, respectively, by jointly optimizing the transmit coefficients at the WDs and the receive beamforming vector at the AP over different subcarriers, subject to the maximum transmit power constraints at individual WDs. However, both problems are shown to be non-convex due to the coupling of the transmit coefficients and the receive beamforming vectors. We address the two problems by considering the special case with a single receive antenna and the general case with multiple receive antennas at the AP, respectively. The main results of this work are summarized as follows.
\begin{itemize}
	\item
	For the special single-input single-output (SISO) case with a single antenna at the AP, we employ the Lagrange-duality method to obtain semi-closed-form globally optimal solutions to both problems. It is shown that at each subcarrier, the optimized power control policy at the WDs follows a regularized channel inversion structure for average MSE minimization and an on-off regularized channel inversion for computation outage probability minimization, where the regularization at each WD is dependent on its transmit power budget and channel estimation error. Notably, a non-zero computation MSE is inevitable at each subcarrier due to channel estimation errors even when the transmit power at each WD goes to infinity.
	\item
	For the general SIMO case with multiple antennas at the AP, we propose efficient algorithms based on alternating optimization and convex optimization to obtain converged solutions to both problems, where the transmit coefficients and the receive beamforming vectors are updated alternately with the other given. It is observed that the optimized power control at WDs follows (on-off) regularized channel inversion structures, similar to the SISO case; while the optimized receive beamforming at the AP follows a sum-minimum MSE (MMSE) structure, which is to better aggregate signals from all the WDs to facilitate the functional computation. Furthermore, it is shown that with massive receive antennas, the computation MSE at each subcarrier would vanish when the channel vectors are i.i.d., thus showing the benefit of using massive antennas to mitigate channel estimation errors in AirComp. 	
	\item
	Finally, numerical results are provided to showcase the impact of channel estimation errors on the computation MSE. 
	The effectiveness of the proposed designs is validated as compared to the benchmark scheme without considering the CSI error, and those with equal power allocation and channel inversion power control over different subcarriers.
\end{itemize}

It is worth noting that there have been some prior works studying AirComp \cite{wu2019computation} and Air-FEEL \cite{zhu2019broadband, zhu2021one, sahin2021distributed, shao2022federated} over OFDM systems. The authors in \cite{wu2019computation} analyzed the computation rates for coded AirComp over SISO OFDM systems. 
The authors in \cite{zhu2019broadband} exploited the uncoded OFDM AirComp over SISO channels (versus SIMO in this work) for Air-FEEL, in which the suboptimal channel inversion power control (versus the optimized power control in this work) was considered. \cite{zhu2021one} and \cite{sahin2021distributed} studied the sign stochastic gradient descent (signSGD) for Air-FEEL over OFDM systems by considering limited constellation sizes such as binary phase shift keying (BPSK) modulation. Furthermore, \cite{shao2022federated} investigated the uncoded Air-FEEL estimator design at the AP for SISO OFDM systems in the presence of channel gain mismatch and synchronization errors. Different from these prior works \cite{wu2019computation, zhu2019broadband, zhu2021one, sahin2021distributed, shao2022federated}, this paper considers the more general SIMO setup with uncoded transmit signals, in which the joint power control and receive beamforming design is developed by taking into account the imperfect CSI caused by channel estimation errors.

The remainder of the paper is organized as follows. Section II presents the uncoded AirComp model in SIMO OFDM systems with imperfect CSI, and formulates the average MSE minimization problem for best-effort computation tasks as well as the computation outage probability minimization problem for error-constrained computation tasks. Section III presents the optimal transceiver design solutions to the two formulated problems in the special SISO case. Section IV proposes the optimized transceiver designs for the two formulated problems in the general SIMO case. Finally, Section V presents numerical results, followed by the conclusion in Section VI.

\textit{Notations}: Bold lower-case and upper-case letters are used for vectors and matrices, respectively. For a vector \(\boldsymbol{a}\), symbols \(\boldsymbol{a}^\dag\), \(\boldsymbol{a}^H\), and \(\|\boldsymbol{a}\|\) denote its conjugate, conjugate transpose, and Euclidean norm, respectively. \(\boldsymbol{I}\) denotes the identity matrix whose dimension will be clear from the context. \(\mathbb{C}^{m\times n}\) denotes the space of \(m \times n\) complex matrices. \(\mathbb{E}[\cdot]\) denotes the statistic expectation. \(\mathcal{CN}(\boldsymbol{\mu}, \boldsymbol{R})\) denotes the circularly symmetric complex Gaussian (CSCG) distribution with mean vector \(\boldsymbol{\mu}\) and covariance matrix \(\boldsymbol{R}\). \(\lceil \cdot \rceil\) denotes the function of rounding up to an integer.

\section{System Model and Problem Formulation}

\subsection{AirComp over OFDM systems}

\begin{figure}[tb]
	\centering
	{\includegraphics[width=0.35\textwidth]{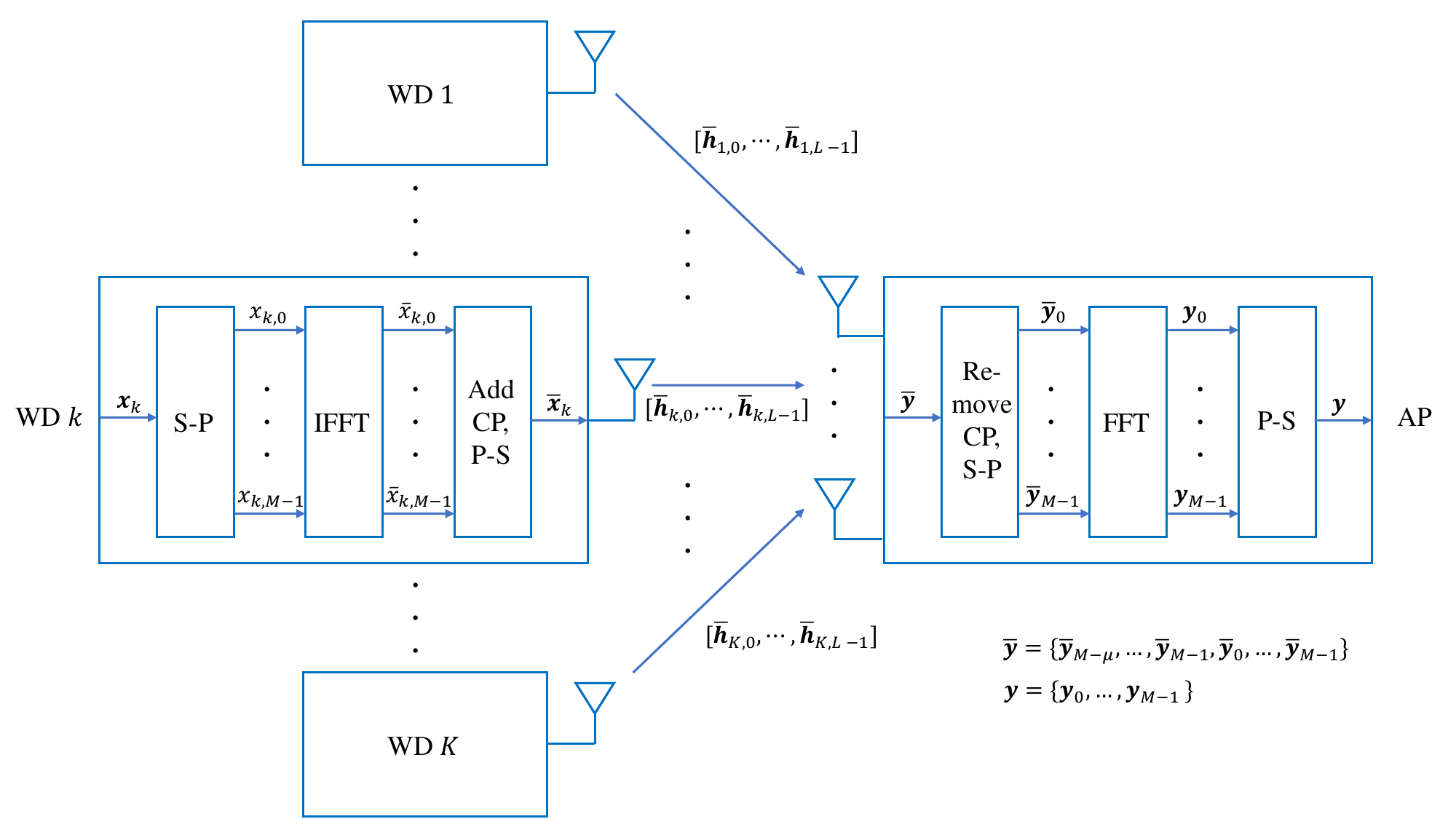}}
	\caption{The uncoded AirComp in OFDM system (S-P: serial-to-parallel converter; P-S: parallel-to-serial converter).}
\end{figure}
We consider the uncoded AirComp implemented over an OFDM system, as illustrated in Fig. 1. In this system, an AP aims to compute the function values of distributed data from a set \(\mathcal{K} \triangleq \{1,\dots, K\}\) of \(K \ge 1\) WDs. Each WD is equipped with a single transmit antenna, while the AP is equipped with \(N_r \ge 1\) receive antennas. Let \(M \ge 1\) denote the number of subcarriers in the OFDM system and \(\mathcal{M} \triangleq \{0,\dots, M-1\}\) denote the set of subcarriers. Each OFDM symbol consists of a set \(\mathcal{M}\) of \(M\) data samples and a set \(\mathcal{U} \triangleq \{1, \dots, \mu\}\) of \(\mu\) samples for the cyclic prefix (CP).

In the OFDM-AirComp system, the AP aims to compute \(M\) function values over an OFDM symbol, with each function value corresponding to one subcarrier. Let \(s_{k,m}\) denote the transmitted message over subcarrier \(m \in \mathcal{M}\) at WD \(k \in \mathcal{K}\). Here, \(s_{k,m}\)'s are assumed to be independent complex random variables, each with zero mean and unit variance.\footnote{In this paper, we consider independent sources (with \(\mathbb{E}\big[s_{i,m}^2\big] = 1\) and \(\mathbb{E}\big[s_{i,m} s_{j,m}\big] = 0, \forall m\in\mathcal M, i,j \in \mathcal{K}, i \ne j\)) for the ease of analysis, similar to prior works on AirComp \cite{zhu2018mimo}.} The objective of the AP is to compute the averaging function of the transmitted messages from all the \(K\) WDs for each subcarrier \(m\),\footnote{Our proposed designs can be extended to other nomographic functions, such as geometric mean, polynomial, and Euclidean norm, by employing appropriate pre-processing at the WDs and post-processing at the AP \cite{goldenbaum2014nomographic}.} which is expressed as
\begin{equation} \label{f_m}
	f_m = \frac{1}{K} \sum_{k \in \mathcal{K}} s_{k,m}, \forall m \in \mathcal{M}.
\end{equation}

First, we consider the signal transmission at each WD \(k \in \mathcal{K}\). Let \(b_{k,m}\) denote the transmit coefficient over subcarrier \(m \in \mathcal{M}\) at WD \(k\). Accordingly, the transmitted signal over the \(M\) subcarriers at WD \(k\) is given by \(\boldsymbol{x}_k = \big[
	x_{k,0}, \dots, x_{k,M-1}
\big]^T\), where \(x_{k,m} =  b_{k,m} s_{k,m}\).\footnote{Notice that in this system, the transmitted signal at each subcarrier is viewed as a continuous signal instead of a discrete one. Although the signal needs to be quantized in such digital communication systems, we can approximate it as a continuous one provided that the quantization levels are sufficiently high.} Let \(P_k\) denote the transmit power budget at WD \(k\). Accordingly, we have the constraint as follows.
\begin{equation}
	\sum_{m\in\mathcal{M}} \mathbb{E}\big[|x_{k,m}|^2\big] = \sum_{m\in\mathcal{M}} |b_{k,m}|^2 \le P_k, \forall k \in \mathcal{K}.
\end{equation}
For each WD \(k\), let \(\bar{\boldsymbol{x}}_k \in \mathbb{C}^{(M+L) \times 1}\) denote the time-domain transmitted signal vector over one OFDM symbol, consisting of \(M+\mu\) samples for both data and CP, which is given by
\begin{equation}
	\bar{\boldsymbol{x}}_k = \big[
		\bar{x}_{k,M-\mu}, \dots, \bar{x}_{k,M-1}, \bar{x}_{k,0}, \dots, \bar{x}_{k,M-1}
	\big]^T, \forall k \in \mathcal{K}.
\end{equation}
As shown in Fig. 1, WD \(k\) implements the inverse discrete Fourier transform (IDFT) or inverse fast Fourier transform (IFFT) on \(\boldsymbol{x}_k\) to obtain \(\bar{\boldsymbol{x}}_k\). Thus, the transmitted signal at sample \(n\) by WD \(k\) is given by
\begin{equation}
	\bar{x}_{k,n} = \frac{1}{\sqrt{M}} \sum_{m \in \mathcal{M}} x_{k,m} e^{j2\pi\frac{mn}{M}}, \forall k \in \mathcal{K}, n \in \mathcal{M}.
\end{equation}

Next, we consider the wireless channels from the WDs to the AP. Suppose that the distance from WD \(k \in \mathcal{K}\) to the AP is denoted by \(d_k\), and the corresponding transmission delay of the line-of-sight (LoS) path is \(\alpha_k = \frac{d_k}{c}\), where \(c\) denotes the speed of light. To enable time synchronization for facilitating AirComp, each WD \(k\) implements a timing advance \(\beta_k\) based on its measured transmission delay. In practice, due to inaccurate time synchronization, the timing advance \(\beta_k\) may deviate slightly from \(\alpha_k\). In this case, we assume that the reference time at the AP is chosen based on the first (or the LoS) signal path from all the \(K\) WDs. Consequently, the timing difference at each WD \(k\) can be characterized by the relative delay given by 
\begin{equation}
	\delta_k = \alpha_k - \beta_k - \min_{i \in \mathcal{K}}(\alpha_i-\beta_i), \forall k \in \mathcal{K}.
\end{equation}
Furthermore, we consider the frequency-selective fading channel from each WD \(k\) to the AP, where the delay spread is denoted by  \(\tau_k\). By combining the relative delay \(\delta_k\) and the delay spread \(\tau_k\), we represent the multi-tap SIMO channel from WD \(k\) to the AP as \(\{
	\bar{\boldsymbol{h}}_{k,0}, \dots, \bar{\boldsymbol{h}}_{k,L-1}
\}\), where \(L = \max_{k\in\mathcal{K}} \lceil \frac{\tau_k+\delta_k}{T} \rceil\) denotes the number of taps, with \(T\) denoting the symbol period and \(\bar{\boldsymbol{h}}_{k,l} \in \mathbb{C}^{N_r \times 1}\) denoting the channel vector of the \(l\)-th tap, \(\forall l \in \mathcal{L} \triangleq \{0, \dots, L-1\}\).\footnote{Notice that \(\mu \ge L\) is sufficient to mitigate the inter-symbol interference (ISI) among different OFDM blocks. This indicates that a coarse time synchronization with \(\delta_k < \mu T, \forall k \in \mathcal{K}\) is adequate for implementing AirComp over OFDM systems.} Notice that for the first \(\Delta_k \triangleq \lceil \frac{\delta_k}{T} \rceil\) taps caused by the timing difference, we have \(\bar{\boldsymbol{h}}_{k,l} = \boldsymbol{0}, \forall l \in \{0, \dots, \Delta_k - 1\}, k \in \mathcal{K}\), for notational convenience.

Then, we consider the received signal at the AP based on the transmitted signals from WDs and the channel responses. For data sample \(n\), the received signal at the AP receiver is given by
\begin{equation}
	\bar{\boldsymbol{y}}_n = \sum_{k \in \mathcal{K}} \sum_{l \in \mathcal{L}} \bar{\boldsymbol{h}}_{k,l} \bar{x}_{k,n-l} + \bar{\boldsymbol{z}}_n, \forall n \in \mathcal{M},
\end{equation}
where \(\bar{\boldsymbol{z}}_n \sim \mathcal{CN}(\boldsymbol{0}, \sigma_z^2 \boldsymbol{I})\) denotes the additive white Gaussian noise (AWGN) at the AP receiver. By performing the discrete Fourier transform (DFT) or fast Fourier transform (FFT), the received signal at each subcarrier \(m\) is expressed as
\begin{equation}
	\boldsymbol{y}_m = \sum_{k \in \mathcal{K}} \boldsymbol{h}_{k,m} b_{k,m} s_{k,m} + \boldsymbol{z}_m, \forall m \in \mathcal{M},
\end{equation}
where \(\boldsymbol{z}_m = \frac{1}{\sqrt{M}} \sum_{n \in \mathcal{M}} \bar{\boldsymbol{z}}_n e^{-j2\pi\frac{mn}{M}} \sim \mathcal{CN}(\boldsymbol{0}, \sigma_z^2 \boldsymbol{I})\) denotes the noise vector at each subcarrier \(m\) and \(\boldsymbol{h}_{k,m} = \sum_{l \in \mathcal{L}} \bar{\boldsymbol{h}}_{k,l} e^{-j2\pi\frac{ml}{M}}\) denotes the channel vector from WD \(k \in \mathcal{K}\) to the AP over subcarrier \(m \in \mathcal{M}\).

In practice, the AP adopts a receive beamforming vector \(\boldsymbol{w}_m \in \mathbb{C}^{N_r \times 1}\) for data aggregation at each subcarrier \(m\). By further performing the averaging operation, the recovered average function value over subcarrier \(m\) at the AP is expressed as
\begin{equation} \label{fh}
	\hat{f}_m = \frac{1}{K} \boldsymbol{w}_m^H \boldsymbol{y}_m, \forall m \in \mathcal{M}.
\end{equation}

\subsection{Computation MSE Minimization with Imperfect CSI}

Next, we present the computation MSE between the ground truth value \(f_m\) in \eqref{f_m} and the recovered value \(\hat{f}_m\) in \eqref{fh} taking into account practical channel estimation errors. We consider TDD systems, where the WDs and the AP estimate the associated channel vectors based on the reverse links by exploiting the uplink-downlink channel reciprocity. This process, however, induces the channel estimation errors. Let \(\hat{\boldsymbol{h}}_{k,m} \in \mathbb{C}^{N_r \times 1}\) denote the estimated channel vector of WD \(k\) at subcarrier \(m\). Consequently, we have
\begin{equation}
	\hat{\boldsymbol{h}}_{k,m} = \boldsymbol{h}_{k,m} + \boldsymbol{e}_{k,m}, \forall k \in \mathcal{K}, \forall m \in \mathcal{M},
\end{equation}
where \(\boldsymbol{e}_{k,m} \sim \mathcal{CN}(\boldsymbol{0}, \sigma_{e,k}^2 \boldsymbol{I})\) denotes the channel estimation error for estimating \(\boldsymbol{h}_{k,m}\).

The receive beamforming vector \(\boldsymbol{w}_m\) at each subcarrier \(m\) is designed based on the estimated CSI \(\hat{\boldsymbol{h}}_{k,m}\)'s. Accordingly, the recovered signal in \eqref{fh} is reexpressed as
\begin{equation} \label{fhh}
	\hat{f}_m = \frac{1}{K} \sum_{k \in \mathcal{K}} \boldsymbol{w}_m^H (\hat{\boldsymbol{h}}_{k,m} - \boldsymbol{e}_{k,m}) b_{k,m} s_{k,m} + \boldsymbol{w}_m^H \boldsymbol{z}_m, \forall m \in \mathcal{M}.
\end{equation}

We utilize the computation MSE at each subcarrier \(m\) as the performance metric of AirComp, which is expressed as follows to characterize the distortion between the recovered value \(\hat{f}_m\) in \eqref{fhh} and the ground truth average \(f_m\) in \eqref{f_m}, i.e.,
\begin{equation} \label{mse}
	\begin{aligned}
		\mathrm{MSE}_m &\big(\{b_{k,m}\}, \boldsymbol{w}_m\big) = \mathbb{E} \big[|\hat{f}_m-f_m|^2\big] \\
		=& \frac{1}{K^2} \Big(\underbrace{\sum_{k \in \mathcal{K}} |\boldsymbol{w}_m^H \hat{\boldsymbol{h}}_{k,m} b_{k,m} - 1|^2}_{\textrm{Signal misalignment error}} + \underbrace{\|\boldsymbol{w}_m\|^2 \sigma_z^2}_{\textrm{Noise-induced error}} \\
		&+ \underbrace{\sum_{k \in \mathcal{K}} \|\boldsymbol{w}_m\|^2 \sigma_{e,k}^2 |b_{k,m}|^2}_{\textrm{CSI-related error}}\Big), \forall m \in \mathcal{M}.
	\end{aligned}
\end{equation}
Notice that in \eqref{mse}, the expectation is taken over the randomness of \(\{s_{k,m}\}\), \(\{\boldsymbol{e}_{k,m}\}\), and \(\boldsymbol{z}_m\), while the last equality holds as the sources \(s_{k,m}\)'s are independent. For each subcarrier \(m\), the MSE in \eqref{mse} consists of three terms: the signal misalignment error, the noise-induced error, and the CSI-related error (due to channel estimation errors). This is different from prior studies with perfect CSI (e.g., \cite{cao2020optimized}), where only the first two terms are present.

In particular, we consider two scenarios with different computation tasks. In the first scenario with best-effort computation tasks, the AP aims to compute a certain number of function values over a specified time duration, and we are interested in the computation errors of the batch of functions across all the \(M\) subcarriers as a whole. Therefore, we employ the average computation MSE over all the \(M\) subcarriers as the performance metric, i.e, \(\mathrm{MSE}_{\mathrm{avg}}\big(\{b_{k,m}\}, \{\boldsymbol{w}_m\}\big) \triangleq \frac{1}{M} \sum_{m=1}^{M} \mathrm{MSE}_m\big(\{b_{k,m}\}, \boldsymbol{w}_m\big)\).
In the second scenario with error-constrained computation, each function value is utilized in mission-critical applications, and we need to ensure the individual computation error of each function at each subcarrier. In this case, we define the computation outage or unsuccessful computation for each subcarrier \(m\) as the event when computation MSE \(\mathrm{MSE}_m\big(\{b_{k,m}\}, \boldsymbol{w}_m\big)\) exceeds a given threshold \(\Gamma\). Accordingly, we define the outage indicator function for each subcarrier \(m \in \mathcal{M}\) as
\begin{equation}
	\mathcal{I}_m\big(\{b_{k,m}\}, \boldsymbol{w}_m\big) =
	\begin{cases}
		0, \ \text{if} \ \mathrm{MSE}_m\big(\{b_{k,m}\}, \boldsymbol{w}_m\big) \le \Gamma, \\
		1, \ \text{otherwise}.
	\end{cases}
\end{equation}
Accordingly, we utilize the computation outage probability or the percentage of unsuccessfully computed functions over the \(M\) subcarriers as the performance metric, which is defined as \(\mathrm{MSE}_{\mathrm{out}}\big(\{b_{k,m}\}, \{\boldsymbol{w}_m\}\big) \triangleq \frac{1}{M} \sum_{m \in \mathcal{M}} \mathcal{I}_m\big(\{b_{k,m}\}, \boldsymbol{w}_m\big)\).

For the two scenarios, we aim to minimize the average computation MSE and the computation outage probability, respectively, by jointly optimizing the transmit coefficients \(\{b_{k,m}\}\) at the WDs and the receive beamforming vector \(\{\boldsymbol{w}_m\}\) at the AP, subject to the individual power budgets at the WDs. The two problems are formulated as (P1) and (P2) as follows, respectively.
\begin{equation*}
	\begin{aligned}
		(\text{P1}): \min_{\{b_{k,m}\}, \{\boldsymbol{w}_m\}} &\
		\frac{1}{M} \sum_{m=1}^{M} \mathrm{MSE}_m\big(\{b_{k,m}\}, \boldsymbol{w}_m\big) \\
		\mathrm{s.t.} &\ \sum_{m \in \mathcal{M}} |b_{k,m}|^2 \le P_k, \forall k \in \mathcal{K}.
	\end{aligned}
\end{equation*}
\begin{equation*}
	\begin{aligned}
		(\text{P2}): \min_{\{b_{k,m}\}, \{\boldsymbol{w}_m\}} &\ \frac{1}{M} \sum_{m \in \mathcal{M}} \mathcal{I}_m\big(\{b_{k,m}\}, \boldsymbol{w}_m\big) \\
		\mathrm{s.t.} &\ \sum_{m \in \mathcal{M}} |b_{k,m}|^2 \le P_k, \forall k \in \mathcal{K}.
	\end{aligned}
\end{equation*}

Notice that both problems (P1) and (P2) are non-convex due to the coupling between \(\{b_{k,m}\}\) and \(\{\boldsymbol{w}_m\}\) in the objective function. As a result, they are challenging to be optimally solved. In the following, we will first present the globally optimal solutions to the two problems for the special SISO case with \(N_r = 1\). Afterward, we will develop efficient algorithms to find high-quality though suboptimal solutions for the general SIMO case with \(N_r > 1\).

\section{Optimal Solutions to Problems (P1) and (P2) with \(N_r = 1\)}

This section presents the optimal solutions to problems (P1) and (P2) in the special SISO case with \(N_r = 1\). In this case, the channel vectors \(\hat{\boldsymbol{h}}_{k,m}\)'s are reduced to scalars \(\hat{h}_{k,m}\)'s. Without loss of optimality, we express the receive beamforming vectors \(\{\boldsymbol{w}_m\}\) as real receive denoising factors \(\{w_m\}\), and set the transmit coefficients as \(b_{k,m} = \tilde{b}_{k,m} \frac{\hat{h}_{k,m}^\dag}{|\hat{h}_{k,m}|}, \forall k \in \mathcal{K}, m \in \mathcal{M}\), in order to align the received signal phases for minimizing the MSE at each subcarrier \(m\), where \(\tilde{b}_{k,m} \ge 0\) denotes the transmit amplitude at WD \(k\). 
As such, problems (P1) and (P2) in the SISO case can be reformulated as problems (P1.1) and (P2.1) as follows, in which the constant coefficients \(\frac{1}{K^2}\) and \(\frac{1}{M}\) are dropped for notational convenience.
\begin{subequations}
	\begin{align}
		(\text{P1.1}): &\min_{\{\tilde{b}_{k,m} \ge 0\}, \{w_m\}} \
		\sum_{m \in \mathcal{M}} \Big(\sum_{k \in \mathcal{K}} \big((w_m |\hat{h}_{k,m}| \tilde{b}_{k,m} - 1)^2 \nonumber \\
		 +& w_m^2 \sigma_{e,k}^2 \tilde{b}_{k,m}^2 \big) + w_m^2 \sigma_z^2 \Big) \triangleq \sum_{m \in \mathcal{M}} \widetilde{\mathrm{MSE}}_m\big(\{\tilde{b}_{k,m}\}, w_m\big) \label{O1.1} \\
		\mathrm{s.t.} &\ \sum_{m \in \mathcal{M}} \tilde{b}_{k,m}^2 \le P_k, \forall m \in \mathcal{M}. \label{C1.1}
	\end{align}
\end{subequations}
\begin{subequations}
	\begin{align}
		(\text{P2.1}): \min_{\{\tilde{b}_{k,m} \ge 0\}, \{w_m\}} &\ \sum_{m \in \mathcal{M}} \tilde{\mathcal{I}}_m\big(\{\tilde{b}_{k,m}\}, w_m\big) \\
		\mathrm{s.t.} &\ \sum_{m \in \mathcal{M}} \tilde{b}_{k,m}^2 \le P_k, \forall k \in \mathcal{K}, \label{C2.1}
	\end{align}
\end{subequations}
where the outage indicator function at each subcarrier \(m \in \mathcal{M}\) becomes
\begin{equation}
	\tilde{\mathcal{I}}_m\big(\{\tilde{b}_{k,m}\}, w_m\big) =
	\begin{cases}
		0, \ \text{if} \ \widetilde{\mathrm{MSE}}_m \big(\{\tilde{b}_{k,m}\}, w_m\big) \le K^2 \Gamma, \\
		1, \ \text{otherwise}.
	\end{cases}
\end{equation}
Although (P1.1) and (P2.1) are not convex, they satisfy the time-sharing condition when \(M\) becomes sufficiently large, in which case the strong duality holds between primal problem (P1.1) or (P2.1) and its corresponding dual problem \cite{wei2006dual}. In the following, we utilize the Lagrange-duality method to obtain the optimal solutions to (P1.1) and (P2.1) for scenarios with best-effort and error-constrained computation tasks, respectively.

\subsection{Optimal Solution to Average MSE Minimization Problem (P1.1)}

This subsection presents the optimal solution to problem (P1.1) for the best-effort computation scenario. Let \(\mu_k \ge 0\) denote the dual variable associated with the transmit power constraint in \eqref{C1.1} for each WD \(k \in \mathcal{K}\). The Lagrangian of (P1.1) is given by \(\mathcal{L}_1 \big(\{\tilde{b}_{k,m}\}, \{w_m\}, \{\mu_k\}\big) = \sum_{m \in \mathcal{M}} \sum_{k \in \mathcal{K}} \big((w_m |\hat{h}_{k,m}| \tilde{b}_{k,m} - 1)^2 + w_m^2 \sigma_{e,k}^2 \tilde{b}_{k,m}^2\big) + \sum_{m \in \mathcal{M}} w_m^2 \sigma_z^2 + \sum_{k \in \mathcal{K}} \mu_k(\sum_{m \in \mathcal{M}} \tilde{b}_{k,m}^2 - P_k)\).
Accordingly, the dual function of (P1.1) is
\begin{equation} \label{dual1}
	g_1\big(\{\mu_k\}\big) = \min_{\{\tilde{b}_{k,m}\}, \{w_m\}} \ \mathcal{L}_1\big(\{\tilde{b}_{k,m}\}, \{w_m\}, \{\mu_k\}\big),
\end{equation}
and the dual problem of (P1.1) is
\begin{equation*}
	(\text{D1.1}): \max_{\{\mu_k\}} \ g_1\big(\{\mu_k\}\big), \ \mathrm{s.t.} \ \mu_k \ge 0, \forall k \in \mathcal{K}.
\end{equation*}

Since the strong duality holds between (P1.1) and (D1.1), we can solve problem (P1.1) by equivalently solving problem (D1.1) \cite{boyd2004convex}. We then have the following proposition. 

\textit{Proposition 1}: Suppose that \(\{\mu_k^{\mathrm{I}} \ge 0\}\) denotes the optimal dual solution to problem (D1.1). The optimal solution to problem (P1.1) is given by \(\{\tilde{b}_{k,m}^{\mathrm{I}}\}\) and \(\{w_m^{\mathrm{I}}\}\), where
\begin{equation} \label{b^opt}
	\tilde{b}_{k,m}^{\mathrm{I}} = \frac{w_m^{\mathrm{I}} |\hat{h}_{k,m}|}{(w_m^{\mathrm{I}})^2 (|\hat{h}_{k,m}|^2 + \sigma_{e,k}^2) + \mu_k^{\mathrm{I}}}, \forall k \in \mathcal{K}, m \in \mathcal{M},
\end{equation}
and \(\{w_m^{\mathrm{I}}\}\) can be obtained via bisection search based on the following equalities. 
\begin{equation} \label{w^opt}
	\sum_{k \in \mathcal{K}} \frac{|\hat{h}_{k,m}|^2 \mu_k}{\big((w_m^{\mathrm{I}})^2 (|\hat{h}_{k,m}|^2 + \sigma_{e,k}^2) + \mu_k^{\mathrm{I}}\big)^2} = \sigma_z^2, \forall m \in \mathcal{M}.
\end{equation}

\textit{Proof}: See Appendix A.
\hfill \(\square\)

\textit{Remark 1}: Proposition 1 provides interesting insights into the optimal solution structure. It is observed in \eqref{b^opt} that the optimal transmit amplitude solution (or power control policy) \(\{\tilde{b}_{k,m}^{\mathrm{I}}\}\) follows a regularized channel inversion structure, where the regularization term \((w_m^{\mathrm{I}})^2 \sigma_{e,k}^2 + \mu_k^{\mathrm{I}}\) depends on the channel estimation error \(\sigma_{e,k}^2\) and the transmit power budget. To be specific, based on the complementary slackness condition that \(\mu_k^{\mathrm{I}} \big(\sum_{m \in \mathcal{M}} (\tilde{b}_{k,m}^{\mathrm{I}})^2 - P_k\big) = 0\), it follows that for each WD \(k\), if the transmit power constraint is inactive at the optimality, then we have \(\mu_k^{\mathrm{I}} = 0\). Accordingly, the transmit coefficients are given by \(\tilde{b}_{k,m}^{\mathrm{I}} = \frac{|\hat{h}_{k,m}|}{w_m^{\mathrm{I}} (|\hat{h}_{k,m}|^2 + \sigma_{e,k}^2)}, \forall m \in \mathcal{M}\), where the regularization only depends on the channel estimation error \(\sigma_{e,k}^2\). By contrast, for each WD \(k\) with \(\mu_k^{\mathrm{I}} > 0\), the transmit power constraint is active at the optimality and the transmit coefficients are given by \eqref{b^opt}, where the regularization depends on both \(\sigma_{e,k}^2\) and \(\mu_k^{\mathrm{I}}\). 

It is also interesting to analyze the average computation MSE when each WD has asymptotically high transmit power (i.e., \( P_k \to \infty, \forall k\in\mathcal K\)), for which we have the following proposition.

\textit{Proposition 2}: When \(P_k \to \infty, \forall k \in \mathcal{K}\), it follows that \(\mathrm{MSE}_m \to \Gamma_{m,1} \triangleq \frac{1}{K^2} \sum_{k \in \mathcal{K}} \frac{\sigma_{e,k}^2}{|\hat{h}_{k,m}|^2 + \sigma_{e,k}^2}\), \(\forall m \in \mathcal{M}\). Accordingly, we have \(\mathrm{MSE}_{\mathrm{avg}} \to \frac{1}{M} \sum_{m \in \mathcal{M}} \Gamma_{m,1}\). 

\textit{Proof}: 
When \(P_k \to \infty, \forall k \in \mathcal{K}\), it can be shown that the optimal receive denoising factors and the transmit amplitudes are given by \(w_m^{\mathrm{I}} \to 0\) and \(\tilde{b}_{k,m}^{\mathrm{I}} = \frac{|\hat{h}_{k,m}|}{w_m^{\mathrm{I}} (|\hat{h}_{k,m}|^2 + \sigma_{e,k}^2)}\), respectively, \(\forall k \in \mathcal{K}, m \in \mathcal{M}\). By substituting them into \eqref{O1.1}, we have \(\mathrm{MSE}_m \to \Gamma_{m,1}, \forall m \in \mathcal{M}\). 
\hfill \(\square\)

Proposition 2 reveals that due to the existence of channel estimation errors \(\sigma_{e,k}^2\)'s, a non-zero average MSE lower bound becomes inevitable even when the transmit powers at WDs approach infinity. This is different from the case with perfect CSI (i.e., \(\sigma_{e,k}^2 = 0, \forall k \in \mathcal{K}\)), in which \(\mathrm{MSE}_{\mathrm{avg}} \to 0\) when \(P_k \to \infty, \forall k \in \mathcal{K}\).

\subsection{Optimal Solution to Computation Outage Probability Minimization Problem (P2.1)}

In this subsection, we present the optimal solution to problem (P2.1) for the error-constrained computation scenario. 
Notice that at each subcarrier \(m\), if \(\Gamma\) is less than or equal to the minimum MSE \(\Gamma_{m,1}\) achieved when the transmit powers tend to infinity (as described in Proposition 2), then it always holds that \(\tilde{\mathcal{I}}_m\big(\{\tilde{b}_{k,m}\}, w_m\big) = 1\). By contrast, if \(\Gamma \ge \frac{1}{K}\), then we can achieve \(\tilde{\mathcal{I}}_m\big(\{\tilde{b}_{k,m}\}, w_m\big) = 0\) by even setting \(\tilde{b}_{k,m} = 0, \forall k \in \mathcal K\). For these two trivial cases, we straightforwardly have \(\{\tilde{b}_{k,m} = 0\}\) and \(w_m = 0\) to be the optimal solution. In the following, we focus on the non-trivial case with \(\Gamma_{m,1} < \Gamma < \frac{1}{K}, \forall m \in \mathcal{M}\).

We employ the Lagrange-duality method to optimally solve problem (P2.1). Let \(\mu_k \ge 0\) denote the dual variable associated with the transmit power constraint in \eqref{C2.1} for WD \(k \in \mathcal{K}\). The Lagrangian of (P2.1) is given by \(\mathcal{L}_2\big(\{\tilde{b}_{k,m}\}, \{w_m\}, \{\mu_k\}\big) = \sum_{m \in \mathcal{M}} \tilde{\mathcal{I}}_m\big(\{\tilde{b}_{k,m}\}, w_m\big) + \sum_{k \in \mathcal{K}} \mu_k (\sum_{m \in \mathcal{M}} \tilde{b}_{k,m}^2 - P_k)\).
Accordingly, the dual function of (P2.1) is
\begin{equation} \label{dual2}
	g_2\big(\{\mu_k\}\big) = \min_{\{\tilde{b}_{k,m}\}, \{w_m\}} \mathcal{L}_2\big(\{\tilde{b}_{k,m}\}, \{w_m\}, \{\mu_k\}\big),
\end{equation}
and the dual problem of (P2.1) is
\begin{equation*}
	(\text{D2.1}): \max_{\{\mu_k\}} \ g_2\big(\{\mu_k\}\big), \ \mathrm{s.t.} \ \mu_k \ge 0, \forall k \in \mathcal{K}.
\end{equation*}
Since the strong duality holds between problems (P2.1) and (D2.1), we solve (P2.1) by equivalently solving (D2.1) \cite{boyd2004convex}. 

First, we find the dual function \(g_2\big(\{\mu_k\}\big)\) in \eqref{dual2} under given \(\{\mu_k \ge 0\}\). By dropping the constant term \(- \sum_{k \in \mathcal{K}} \mu_k P_k\), the problem in \eqref{dual2} can be decomposed into the following \(M\) subproblems, each corresponding to one subcarrier \(m \in \mathcal{M}\).
\begin{equation*}
	(\text{P3}.m): \min_{\{\tilde{b}_{k,m}\}, w_m} \tilde{\mathcal{I}}_m\big(\{\tilde{b}_{k,m}\}, w_m\big) + \sum_{k \in \mathcal{K}} \mu_k \tilde{b}_{k,m}^2.
\end{equation*}
To solve problem (P3.\(m\)), we need to find the minimum values of \(\tilde{\mathcal{I}}_m\big(\{\tilde{b}_{k,m}\}, w_m\big) + \sum_{k \in \mathcal{K}} \mu_k \tilde{b}_{k,m}^2\) under the cases with \(\tilde{\mathcal{I}}_m\big(\{\tilde{b}_{k,m}\}, w_m\big) = 0\) and \(\tilde{\mathcal{I}}_m\big(\{\tilde{b}_{k,m}\}, w_m\big) = 1\), respectively, and then compare them to find the minimum one. First, in the case with \(\tilde{\mathcal{I}}_m\big(\{\tilde{b}_{k,m}\}, w_m\big) = 1\), i.e., the computation MSE exceeds the threshold \(\Gamma\), the objective of (P3.\(m\)) is minimized to be \(1\) by simply setting \(\{\tilde{b}_{k,m} = 0\}\). Next, in the other case with \(\tilde{\mathcal{I}}_m\big(\{\tilde{b}_{k,m}\}, w_m\big) = 0\), i.e., the computation MSE is no greater than the threshold \(\Gamma\), minimizing the objective of (P3.\(m\)) is equivalent to solving the following problem.
\begin{equation*}
	\begin{aligned}
		(\text{P3.1}.m): \min_{\{\tilde{b}_{k,m}\}, w_m} &\ \sum_{k \in \mathcal{K}} \mu_k \tilde{b}_{k,m}^2 \\
		\mathrm{s.t.} &\ \widetilde{\mathrm{MSE}}_m\big(\{\tilde{b}_{k,m}\}, w_m\big) \le K^2 \Gamma.
	\end{aligned}
\end{equation*}
To facilitate the solution to problem (P3.1.\(m\)), we denote \(\Gamma_{m,2} \triangleq \sum_{k \in \mathcal{K}, \mu_k > 0} \frac{1}{K^2} + \frac{1}{K^2} \sum_{k \in \mathcal{K}, \mu_k = 0} \frac{\sigma_{e,k}^2}{|\hat{h}_{k,m}|^2 + \sigma_{e,k}^2}\) as the minimum MSE for subcarrier \(m\) when the WDs with \(\mu_k = 0\) are switched on with \(P_k \to \infty\) and those with \(\mu_k > 0\) are switched off with \(P_k = 0\). We then have the following lemma, in which only the nontrivial case when at least one \(\mu_k\) is non-zero and \(\Gamma \ne \Gamma_{m,2}\) is considered. 

\textit{Lemma 1}:
Let \(\lambda_m^* > 0\) denote the optimal dual variable associated with the MSE constraint in problem (P3.1.\(m\)). 
The optimal solution to (P3.1.\(m\)) is given by \(\{\tilde{b}_{k,m}^*\}\) and \(w_m^*\), where
\begin{equation} \label{b^out1}
	\begin{aligned}
		\tilde{b}_{k,m}^* = \begin{cases}
			0, \ \text{if} \ \mu_k > 0 \ \text{and} \ \Gamma_{m,2} < \Gamma < \frac{1}{K}, \\
			\frac{w_m^* |\hat{h}_{k,m}|}{(w_m^*)^2 (|\hat{h}_{k,m}|^2 + \sigma_{e,k}^2) + \frac{\mu_k}{\lambda_m^*}}, \ \text{otherwise},
		\end{cases} \forall k \in \mathcal{K},
	\end{aligned}
\end{equation}
and 
\begin{equation} \label{w^out1}
	\begin{aligned}
		w_m^* = \begin{cases}
			\frac{K}{\sigma_z} \sqrt{\Gamma-\Gamma_{m,2}}, \ \text{if} \ \Gamma_{m,2} < \Gamma < \frac{1}{K}, \\
			\frac{1}{\sigma_z} \sqrt{\sum_{k \in \mathcal{K}} \frac{v_m^* |\hat{h}_{k,m}|^2 \mu_k}{(v_m^* |\hat{h}_{k,m}|^2 + v_m^* \sigma_{e,k}^2 + \mu_k)^2}}, \ \text{otherwise}.
		\end{cases}
	\end{aligned}
\end{equation}	
Here, \(v_m^* \triangleq \lambda_m^* (w_m^*)^2 > 0\) is obtained by a bisection search based on
\begin{equation} \label{lam1}
	\sum_{k \in \mathcal{K}} \frac{v_m^* \sigma_{e,k}^2 + \mu_k}{v_m^* |\hat{h}_{k,m}|^2 + v_m^* \sigma_{e,k}^2 + \mu_k} = K^2 \Gamma.
\end{equation}

\textit{Proof}: See Appendix B.
\hfill \(\square\)

Note that in Lemma 1, we first find \(v_m^*\) based on \eqref{lam1}, then obtain \(w_m^*\) by \eqref{w^out1}, and finally determine \(\{\tilde{b}_{k,m}^*\}\) based on \eqref{b^out1} with \(\lambda_m^* = \frac{v_m^*}{(w_m^*)^2}\). Based on Lemma 1, we have the following proposition to solve (P3.\(m\)) and thus problem \eqref{dual2} optimally. Notice that Proposition 3 can be verified by comparing the optimal value of 1 achieved when \(\tilde{\mathcal{I}}_m\big(\{\tilde{b}_{k,m}\}, w_m\big) = 1\) versus that of \(\sum_{k \in \mathcal{K}} \mu_k (\tilde{b}_{k,m}^*)^2\) achieved when \(\tilde{\mathcal{I}}_m\big(\{\tilde{b}_{k,m}\}, w_m\big) = 0\), for which the details are omitted. 

\textit{Proposition 3}: Let \(\{\tilde{b}_{k,m}^\star\}\) and \(w_m^\star\) denote the optimal solution to problem (P3.\(m\)). For each subcarrier \(m\), if \(\sum_{k \in \mathcal{K}} \mu_k (\tilde{b}_{k,m}^*)^2 > 1\), then the optimal solution is \(\{\tilde{b}_{k,m}^\star = 0\}\) and \(w_m^\star = 0\); if \(\sum_{k \in \mathcal{K}} \mu_k (\tilde{b}_{k,m}^*)^2 \le 1\), then the optimal solution is \(\{\tilde{b}_{k,m}^\star = \tilde{b}_{k,m}^*\}\) and \(w_m^\star = w_m^*\).


Next, we solve the dual problem (D2.1), which is convex but not necessarily differentiable. To handle this, we solve (D2.1) by applying subgradient-based methods such as the ellipsoid method \cite{boyd2004convex}. For the objective function \(g_2\big(\{\mu_k\}\big)\), the subgradient at \(\mu_k\) is \(\sum_{m \in \mathcal{M}} (\tilde{b}_{k,m}^\star)^2 - P_k, \forall k \in \mathcal{K}\). By utilizing this subgradient, the ellipsoid method can be implemented efficiently, based on which we can obtain the optimal dual solution to (D2.1) as \(\{\mu_k^{\mathrm{II}}\}\).

Then, we present the optimal solution to the primal problem (P2.1). With the optimal dual variables \(\{\mu_k^{\mathrm{II}}\}\) at hand, the corresponding optimal solutions \(\{\tilde{b}_{k,m}^\star\}\) and \(\{w_m^\star\}\) to problem \eqref{dual2} in Proposition 3 can be directly used for constructing the optimal primal solution to (P2.1), denoted by \(\{\tilde{b}_{k,m}^{\mathrm{II}}\}\) and \(\{w_m^{\mathrm{II}}\}\).

\textit{Remark 2}:
It is observed from \eqref{b^out1} that the optimal transmit amplitude solution (or power control policy) \(\{\tilde{b}_{k,m}^{\mathrm{II}}\}\) follows an on-off regularized channel inversion structure, where the regularization and on-off control are determined by the power budget, channel estimation error, and MSE threshold. Specifically, for each subcarrier \(m\), if \(\sum_{k \in \mathcal{K}} \mu_k^{\mathrm{II}} (\tilde{b}_{k,m}^{\mathrm{II}})^2 > 1\), then the WDs are turned off with \(\tilde{b}_{k,m}^{\mathrm{II}} = 0, \forall k \in \mathcal{K}\), such that the outage happens with \(\tilde{\mathcal{I}}_m\big(\{\tilde{b}_{k,m}\}, w_m\big) = 1\). This is due to the fact that the transmit power required for successful computation is too high in this case. 
By contrast, if \(\sum_{k \in \mathcal{K}} \mu_k^{\mathrm{II}} (\tilde{b}_{k,m}^*)^2 \le 1\), then the computation is successful in this subcarrier with \(\tilde{\mathcal{I}}_m\big(\{\tilde{b}_{k,m}\}, w_m\big) = 0\), for which the transmit policy at the WDs has the following properties.
\begin{itemize}
	\item For WD \(k\), if the transmit power constraint for this WD is inactive or the transmit power is sufficient, then we have \(\mu_k^{\mathrm{II}}= 0\). Accordingly, the optimal transmit amplitude becomes \(\tilde{b}_{k,m}^{\mathrm{II}} = \frac{|\hat{h}_{k,m}|}{w_m^{\mathrm{II}} (|\hat{h}_{k,m}|^2 + \sigma_{e,k}^2)}\), where   the regularization depends solely on \(\sigma_{e,k}^2\), similarly as that for the best-effort computation scenario.
	\item For WD \(k\), if \(\mu_k^{\mathrm{II}} > 0\), then the transmit amplitudes satisfy that \(\sum_{m \in \mathcal{M}} (\tilde{b}_{k,m}^{\mathrm{II}})^2 = P_k\), i.e., the transmit power constraint is active. In this case, WD \(k\) may adopt on-off transmission control depending on the MSE threshold \(\Gamma\). In particular, if \(\Gamma_{m,2} < \Gamma < \frac{1}{K}\), then WD \(k\) is turned off with \(\tilde{b}_{k,m}^{\mathrm{II}} = 0\). This is due to the fact that the transmission of other WDs with sufficient power (especially those with \(\mu_k^{\mathrm{II}} = 0\)) can meet the computation error requirement. On the other hand, if \(\Gamma_{m,1} < \Gamma < \Gamma_{m,2}\), then WD \(k\) transmits with \(\tilde{b}_{k,m}^{\mathrm{II}} = \frac{w_m^{\mathrm{II}} |\hat{h}_{k,m}|}{(w_m^{\mathrm{II}})^2 (|\hat{h}_{k,m}|^2 + \sigma_{e,k}^2) + \frac{\mu_k^{\mathrm{II}}}{\lambda_m^*}}\), where the regularization  depends on \(\sigma_{e,k}^2\), \(\mu_k^{\mathrm{II}}\), and \(\lambda_m^*\). This is different from \eqref{b^opt} for the best-effort computations scenario.
\end{itemize}


It is also interesting to discuss the case when each WD has asymptotically high transmit power (i.e., \(P_k \to \infty, \forall k \in \mathcal{K}\)). In this case, it follows from Proposition 2 that the minimum achievable MSE at each subcarrier \(m\) is given by \(\Gamma_{m,1}\). As a result, if the MSE threshold is set such that \(\Gamma \ge \Gamma_{m,1}, \forall m \in \mathcal{M}\), then the zero computation outage probability \(\mathrm{MSE}_{\mathrm{out}} = 0\) is achievable. However, if the MSE threshold \(\Gamma < \Gamma_{m,1}\) for any subcarrier \(m\), then computation outage becomes inevitable. This is different from the case with perfect CSI, in which \(\mathrm{MSE}_{\mathrm{out}} = 0\) as long as \(P_k \to \infty, \forall k \in \mathcal{K}\).

\section{Proposed Solutions to Problems (P1) and (P2) with \(N_r > 1\)}

This section considers problems (P1) and (P2) in the general SIMO case with \(N_r > 1\). To minimize the computation MSE in this case, with phase alignment, we set the transmit coefficients as \(b_{k,m} = \tilde{b}_{k,m} \frac{\hat{\boldsymbol{h}}_{k,m}^H \boldsymbol{w}_m}{|\boldsymbol{w}_m^H \hat{\boldsymbol{h}}_{k,m}|}, \forall k \in \mathcal{K}, m \in \mathcal{M}\), where \(\tilde{b}_{k,m} \ge 0\) is the transmit amplitude. Therefore, (P1) and (P2) are equivalently reformulated as follows for scenarios with best-effort and error-constrained computation tasks, respectively.
\begin{equation*}
	\begin{aligned}
		(\text{P1.2}): \min_{\{\tilde{b}_{k,m} \ge 0\}, \{\boldsymbol{w}_m\}} &\
		\sum_{m \in \mathcal{M}} \Big(\sum_{k \in \mathcal{K}} \big((|\boldsymbol{w}_m^H \hat{\boldsymbol{h}}_{k,m}| \tilde{b}_{k,m} - 1)^2 \\
		&\ + \|\boldsymbol{w}_m\|^2 \sigma_{e,k}^2 \tilde{b}_{k,m}^2\big) + \|\boldsymbol{w}_m\|^2 \sigma_{z}^2\Big) \\
		&\ \triangleq \sum_{m \in \mathcal{M}} \overline{\mathrm{MSE}}_m\big(\{\tilde{b}_{k,m}\}, \boldsymbol{w}_m\big) \\
		\mathrm{s.t.} &\ \sum_{m \in \mathcal{M}} \tilde{b}_{k,m}^2 \le P_k, \forall k \in \mathcal{K}.
	\end{aligned}
\end{equation*}
\begin{equation*}
	\begin{aligned}
		(\text{P2.2}): \min_{\{\tilde{b}_{k,m} \ge 0\}, \{\boldsymbol{w}_m\}} &\ \sum_{m \in \mathcal{M}} \bar{\mathcal{I}}_m\big(\{\tilde{b}_{k,m}\}, \boldsymbol{w}_m\big) \\
		\mathrm{s.t.} &\ \sum_{m \in \mathcal{M}} \tilde{b}_{k,m}^2 \le P_k, \forall k \in \mathcal{K},
	\end{aligned}
\end{equation*}
where the outage indicator function at each subcarrier \(m \in \mathcal{M}\) becomes
\begin{equation}
	\bar{\mathcal{I}}_m\big(\{\tilde{b}_{k,m}\}, \boldsymbol{w}_m\big) =
	\begin{cases}
		0, \ \text{if} \ \overline{\mathrm{MSE}}_m\big(\{\tilde{b}_{k,m}\}, \boldsymbol{w}_m\big) \le K^2 \Gamma, \\
		1, \ \text{otherwise}.
	\end{cases}
\end{equation}
For problems (P1.2) and (P2.2), to deal with the coupling of the transmit coefficients and the receive combining vectors, we propose efficient algorithms based on alternating optimization, where \(\{\tilde{b}_{k,m}\}\) and \(\{\boldsymbol{w}_m\}\) are updated alternately with the other given.

\subsection{Proposed Solution to Average MSE Minimization Problem (P1.2)}

\subsubsection{Optimal Solution of \(\{\boldsymbol{w}_m\}\) to Problem (P1.2) with Given \(\{\tilde{b}_{k,m}\}\)}

First, we optimize \(\{\boldsymbol{w}_m\}\) in problem (P1.2) under given \(\{b_{k,m}\}\), or equivalently \(\{\tilde{b}_{k,m}\}\). This involves solving the following \(M\) unconstrained subproblems, each for one subcarrier \(m \in \mathcal{M}\).
\begin{equation} \label{1.3}
	\min_{\boldsymbol{w}_m} \ \sum_{k \in \mathcal{K}} (|\boldsymbol{w}_m^H \hat{\boldsymbol{h}}_{k,m} b_{k,m} - 1|^2 + \|\boldsymbol{w}_m\|^2 \sigma_{e,k}^2 \tilde{b}_{k,i}^2) + \|\boldsymbol{w}_m\|^2 \sigma_z^2.
\end{equation}
Problem \eqref{1.3} is convex. Therefore, by setting the gradient of the objective function to zero, the optimal solution to \eqref{1.3} is given by
\begin{equation} \label{w^star}
	\begin{aligned}
		\boldsymbol{w}_m^{\mathrm{III}} =& \Big({\sum_{k \in \mathcal{K}} \tilde{b}_{k,m}^2 (\hat{\boldsymbol{h}}_{k,m} \hat{\boldsymbol{h}}_{k,m}^H + \sigma_{e,k}^2 \boldsymbol{I}) + \sigma_z^2 \boldsymbol{I}}\Big)^{-1} \\
		&\cdot \sum_{k \in \mathcal{K}} \hat{\boldsymbol{h}}_{k,m} b_{k,m}.
	\end{aligned}
\end{equation}

For each subcarrier \(m\), the optimized receive beamforming solution in \eqref{w^star} is observed to have a sum-MMSE structure. This is in order to better aggregate the signals from all the WDs to facilitate the functional computation.

\subsubsection{Optimal Solution of \(\{\tilde{b}_{k,m}\}\) to Problem (P1.2) with Given \(\{\boldsymbol{w}_m\}\)} Next, we optimize \(\{\tilde{b}_{k,m}\}\) for problem (P1.2) under given \(\{\boldsymbol{w}_m\}\), which is equivalently transformed into the following \(K\) subproblems, each for one WD \(k \in \mathcal{K}\).
\begin{equation} \label{1.4}
	\begin{aligned}
		\min_{\{\tilde{b}_{k,m}\}} &\
		\sum_{m \in \mathcal{M}} \big((|\boldsymbol{w}_m^H \hat{\boldsymbol{h}}_{k,m}| \tilde{b}_{k,m} - 1)^2 + \|\boldsymbol{w}_m\|^2 \sigma_{e,k}^2 \tilde{b}_{k,m}^2\big) \\
		\mathrm{s.t.} &\ \sum_{m \in \mathcal{M}} \tilde{b}_{k,m}^2 \le P_k.
	\end{aligned}
\end{equation}
By exploiting the KKT conditions, the optimal solution to problem \eqref{1.4} is given by
\begin{equation} \label{b^star}
	\tilde{b}_{k,m}^{\mathrm{III}} = \frac{|\boldsymbol{w}_m^H \hat{\boldsymbol{h}}_{k,m}|}{|\boldsymbol{w}_m^H \hat{\boldsymbol{h}}_{k,m}|^2 + \|\boldsymbol{w}_m\|^2 \sigma_{e,k}^2 + \mu_k^{\mathrm{III}}}, \forall m \in \mathcal{M},
\end{equation}
where \(\mu_k^{\mathrm{III}} \ge 0\) denotes the optimal dual variable associated with the transmit power constraint in \eqref{1.4}. If \(\sum_{m \in \mathcal{M}} (\frac{|\boldsymbol{w}_m^H \hat{\boldsymbol{h}}_{k,m}|}{|\boldsymbol{w}_m^H \hat{\boldsymbol{h}}_{k,m}|^2 + \|\boldsymbol{w}_m\|^2 \sigma_{e,k}^2})^2 < P_k\), we have \(\mu_k^{\mathrm{III}} = 0\); otherwise, \(\mu_k^{\mathrm{III}}\) is obtained by using a bisection search based on the equality of \(\sum_{m \in \mathcal{M}} (\frac{|\boldsymbol{w}_m^H \hat{\boldsymbol{h}}_{k,m}|}{|\boldsymbol{w}_m^H \hat{\boldsymbol{h}}_{k,m}|^2 + \|\boldsymbol{w}_m\|^2 \sigma_{e,k}^2 + \mu_k^{\mathrm{III}}})^2 = P_k\).

\textit{Remark 3}: The optimized transmit amplitude solution or equivalently power control policy in \eqref{b^star} is observed to exhibit a regularized channel inversion structure (by viewing \(\frac{|\boldsymbol{w}_m^H \hat{\boldsymbol{h}}_{k,m}|^2}{\|\boldsymbol{w}_m\|^2}\) as the equivalent channel power gain), similar to that in \eqref{b^opt} for the SISO case, where the regularization depends on the transmit power budget and channel estimation error. 

\subsubsection{Complete Algorithm for Solving to Problem (P1.2)} The alternating-optimization-based algorithm for solving (P1.2) is implemented in an iterative manner. In each iteration, we first update the transmit coefficients as \(\{\tilde{b}_{k,m}^{\mathrm{III}}\}\) in \eqref{b^star} under given \(\{\boldsymbol{w}_m\}\), and then update the receive beamforming vector as \(\{\boldsymbol{w}_m^{\mathrm{III}}\}\) based on \eqref{w^star} under given \(\{\tilde{b}_{k,m}\}\). Notice that in each iteration, both problems \eqref{1.3} and \eqref{1.4} are optimally solved. Therefore, the updated average MSE is ensured to be monotonically nonincreasing. As the average MSE in (P1.2) is lower bounded, the convergence of our proposed alternating-optimization-based algorithm can be guaranteed. 


It is interesting to discuss the average computation MSE in the cases with sufficient transmit powers or a massive number of receive antennas, for which we have the following propositions.

\textit{Proposition 4}: Under any given receive beamforming vector \(\boldsymbol{w}_m\), if \(P_k \to \infty, \forall k\in\mathcal K\), then we have \(\mathrm{MSE}_m \to \bar{\Gamma}_{m,1} \triangleq \frac{1}{K^2} \sum_{k \in \mathcal{K}} \frac{\|\boldsymbol{w}_m\|^2 \sigma_{e,k}^2}{|\boldsymbol{w}_m^H \hat{\boldsymbol{h}}_{k,m}|^2 + \|\boldsymbol{w}_m\|^2 \sigma_{e,k}^2} \ge \frac{1}{K^2} \sum_{k \in \mathcal{K}} \frac{\sigma_{e,k}^2}{\|\hat{\boldsymbol{h}}_{k,m}\|^2 + \sigma_{e,k}^2}, \forall m \in \mathcal{M}\).
Accordingly, it follows that \(\mathrm{MSE}_{\mathrm{avg}} \to \frac{1}{M} \sum_{m \in \mathcal{M}} \bar{\Gamma}_{m,1}\). 

Proposition 4 can be similarly verified as Proposition 2, where the inequality holds based on the Cauchy-Schwarz inequality. It follows from Proposition 4 that with a finite number of receive antennas, a non-zero MSE is inevitable even when the WDs employ extremely high transmit powers.

\textit{Proposition 5}: 
If \(N_r \to \infty\) and \(\boldsymbol{h}_{k,m}\)'s (and equivalently \(\hat{\boldsymbol{h}}_{k,m}\)'s) are i.i.d. random vectors, then we have \(\mathrm{MSE}_m \to 0, \forall m \in \mathcal{M}\), and accordingly \(\mathrm{MSE}_{\mathrm{avg}} \to 0\). 

\textit{Proof}: In this case, the channel vectors among different WDs become asymptotically orthogonal, i.e., \(\frac{1}{N_r} \hat{\boldsymbol{h}}_{i,m} \hat{\boldsymbol{h}}_{j,m}^H \approx \boldsymbol{0}\) and \(\hat{\boldsymbol{h}}_{i,m} \hat{\boldsymbol{h}}_{i,m}^H \approx N_r \sigma_h^2 \boldsymbol{I}\), \(\forall i,j \in \mathcal{K}, i \neq j, m \in \mathcal{M}\), where \(\sigma_h^2\) denotes the variance of \(\hat{\boldsymbol{h}}_{k,m}\)'s. Accordingly, we have \(\boldsymbol{w}_m^{\mathrm{III}} = \frac{1}{\eta_m^{(1)}+\eta_m^{(2)}+\eta_m^{(3)}} \sum_{k \in \mathcal{K}} \hat{\boldsymbol{h}}_{k,m} b_{k,m}\), where \(\eta_m^{(1)} = \sum_{k \in \mathcal{K}} \tilde{b}_{k,m}^2 N_r \sigma_h^2\), \(\eta_m^{(2)} = \sum_{k \in \mathcal{K}} \tilde{b}_{k,m}^2 \sigma_{e,k}^2\), and \(\eta_m^{(3)} = \sigma_z^2\). By substituting \(\boldsymbol{w}_m^{\mathrm{III}}\) into the objective function of problem (P1.2), the computation MSE becomes \(\mathrm{MSE}_m = \frac{(\eta_m^{(2)}+\eta_m^{(3)})^2 + \eta_m^{(1)}\eta_m^{(2)} + \eta_m^{(1)}\eta_m^{(3)}}{K^2 (\eta_m^{(1)}+\eta_m^{(2)}+\eta_m^{(3)})^2}\). As \(N_r \to \infty\), we have  \(\eta_m^{(1)} \to \infty, \forall m \in \mathcal{M}\). As a result, we have \(\mathrm{MSE}_m \to 0, \forall m \in \mathcal{M}\).
\hfill \(\square\)

Proposition 5 shows that increasing the number of receive antennas is efficient to combat against the imperfect CSI, thus showing the benefit of massive antennas in AirComp.

\subsection{Proposed Solution to Computation Outage Probability Minimization Problem (P2.2)}

\subsubsection{Optimal Solution of \(\{\boldsymbol{w}_m\}\) to Problem (P2.2) with given \(\{\tilde{b}_{k,m}\}\)}

First, we optimize \(\{\boldsymbol{w}_m\}\) for problem (P2.2) under given \(\{\tilde{b}_{k,m}\}\). In this case, we design \(\{\boldsymbol{w}_m\}\) to minimize the computation MSE at each subcarrier. For subcarriers \(m\) with \(\tilde{b}_{k,m} = 0, \forall k \in \mathcal{K}\), there is no need to update \(\boldsymbol{w}_m\). For the other subcarriers, \(\boldsymbol{w}_m\) is optimized as follows, the same as the sum-MMSE beamformer in \eqref{w^star} for problem (P1.2).
\begin{equation} \label{w^out}
	\begin{aligned}
		\boldsymbol{w}_m^{\mathrm{IV}} =& \Big({\sum_{k \in \mathcal{K}} \tilde{b}_{k,m}^2 (\hat{\boldsymbol{h}}_{k,m} \hat{\boldsymbol{h}}_{k,m}^H + \sigma_{e,k}^2 \boldsymbol{I}) + \sigma_z^2 \boldsymbol{I}}\Big)^{-1} \\
		&\cdot \sum_{k \in \mathcal{K}} \hat{\boldsymbol{h}}_{k,m} b_{k,m}.
	\end{aligned}
\end{equation}

\subsubsection{Optimal Solution of \(\{\tilde{b}_{k,m}\}\) to Problem (P2.2) with Given \(\{\boldsymbol{w}_m\}\)}

Next, we optimize \(\{\tilde{b}_{k,m}\}\) for problem (P2.2) under given \(\{\boldsymbol{w}_m\}\), which is expressed as the following problem.
\begin{subequations}
	\begin{align}
		(\text{P2.3}): \min_{\{\tilde{b}_{k,m}\}} &\ \sum_{m \in \mathcal{M}} \bar{\mathcal{I}}_m\big(\{\tilde{b}_{k,m}\}, \boldsymbol{w}_m\big) \\
		\mathrm{s.t.} &\ \sum_{m \in \mathcal{M}} \tilde{b}_{k,m}^2 \le P_k, \forall k \in \mathcal{K}. \label{C2.3}
	\end{align}
\end{subequations}
As problem (P2.3) satisfies the time-sharing condition when \(M\) becomes sufficiently large, we employ the Lagrange-duality method to find its optimal solution.
Before proceeding, notice that similar to problem (P1.2), when \(\Gamma \le \bar{\Gamma}_{m,1} + \frac{\|\boldsymbol{w}_m\|^2 \sigma_z^2}{K^2}\) or \(\Gamma \ge \frac{1}{K} + \frac{\|\boldsymbol{w}_m\|^2 \sigma_z^2}{K^2}\), for any \(m \in\mathcal{M}\) (with \(\bar{\Gamma}_{m,1}\) defined in Proposition 4), we have \(\bar{\mathcal{I}}\big(\{\tilde{b}_{k,m}\}, \boldsymbol{w}_m\big) = 1\) or \(\bar{\mathcal{I}}\big(\{\tilde{b}_{k,m}\}, \boldsymbol{w}_m\big) = 0\). For the two cases, we straightforwardly set \(\tilde{b}_{k,m} = 0, \forall k \in \mathcal{K}\). In the following, we only need to consider the non-trivial case with \(\bar{\Gamma}_{m,1} < \Gamma - \frac{\|\boldsymbol{w}_m\|^2 \sigma_z^2}{K^2} < \frac{1}{K}, \forall m \in \mathcal{M}\).

Let \(\mu_k \ge 0\) denote the dual variable associated with the transmit power constraint in \eqref{C2.3} for WD \(k \in \mathcal{K}\). The Lagrangian of (P2.3) is given by \(\mathcal{L}_3\big(\{\tilde{b}_{k,m}\},\{\mu_k\}\big) = \sum_{m \in \mathcal{M}} \bar{\mathcal{I}}_m\big(\{\tilde{b}_{k,m}\}, \boldsymbol{w}_m\big) + \sum_{k \in \mathcal{K}} \mu_k (\sum_{m \in \mathcal{M}} \tilde{b}_{k,m}^2 - P_k)\).
Accordingly, the dual function of P(2.3) is
\begin{equation} \label{dual3}
	g_3\big(\{\mu_k\}\big) = \min_{\{\tilde{b}_{k,m}\}} \mathcal{L}_3\big(\{\tilde{b}_{k,m}\}, \{\mu_k\}\big),
\end{equation}
and the dual problem of (P2.3) is
\begin{equation*}
	(\text{D2.3}): \max_{\{\mu_k\}} \ g_3\big(\{\mu_k\}\big), \ \mathrm{s.t.} \ \mu_k \ge 0, \forall k \in \mathcal{K}.
\end{equation*}
Since the strong duality holds between problems (P2.3) and (D2.3), we solve (P2.3) by equivalently solving (D2.3) \cite{boyd2004convex}. 

First, we find the dual function \(g_3\big(\{\mu_k\}\big)\) in \eqref{dual3} under given \(\{\mu_k \ge 0\}\). The problem in \eqref{dual3} can be decomposed into the following \(M\) subproblems, each for one subcarrier \(m \in \mathcal{M}\).
\begin{equation*}
	(\text{P4}.m): \min_{\{\tilde{b}_{k,m}\}} \ \bar{\mathcal{I}}_m\big(\{\tilde{b}_{k,m}\}, \boldsymbol{w}_m\big) + \sum_{k \in \mathcal{K}} \mu_k \tilde{b}_{k,m}^2.
\end{equation*}
Similar to problem (P3.\(m\)), we first solve problem (P4.\(m\)) in two cases with \(\bar{\mathcal{I}}_m\big(\{\tilde{b}_{k,m}\}, \boldsymbol{w}_m\big) = 1\) and \(\bar{\mathcal{I}}_m\big(\{\tilde{b}_{k,m}\}, \boldsymbol{w}_m\big) = 0\), respectively, and then compare the optimal values obtained in these two cases. If \(\bar{\mathcal{I}}_m\big(\{\tilde{b}_{k,m}\}, \boldsymbol{w}_m\big) = 1\), then the optimal value of (P4.\(m\)) is 1; if \(\bar{\mathcal{I}}_m\big(\{\tilde{b}_{k,m}\}, \boldsymbol{w}_m\big) = 0\), then (P4.\(m\)) is equivalent to the following problem.
\begin{equation*}
	\begin{aligned}
		(\text{P4.1}.m): \min_{\{\tilde{b}_{k,m}\}} &\ \sum_{k \in \mathcal{K}} \mu_k \tilde{b}_{k,m}^2 \\
		\mathrm{s.t.} &\ \overline{\mathrm{MSE}}_m\big(\{\tilde{b}_{k,m}\}, \boldsymbol{w}_m\big) \le K^2 \Gamma. 
	\end{aligned}
\end{equation*}

To facilitate the solution to problem (P4.1.\(m\)), similar to problem (P3.1.\(m\)), we define \(\bar{\Gamma}_{m,2} \triangleq \sum_{k\in \mathcal{K}, \mu_k^{\mathrm{IV}} > 0} \frac{1}{K^2} + \frac{1}{K^2} \sum_{k \in \mathcal{K}, \mu_k^{\mathrm{IV}} = 0} \frac{\|\boldsymbol{w}_m\|^2 \sigma_{e,k}^2}{|\boldsymbol{w}_m^H \hat{\boldsymbol{h}}_{k,m}|^2 + \|\boldsymbol{w}_m\|^2 \sigma_{e,k}^2}\). We then have the following lemma, in which only the nontrivial case when at least one \(\mu_k\) is non-zero is considered.

\textit{Lemma 2}: 
Let \(\lambda_m^{**} > 0\) denote the optimal dual variable associated with the computation MSE constraint in problem (P4.1.\(m\)). 
The optimal solution to (P4.1.\(m\)) is given by \(\tilde{b}_{k,m}^{**}, \forall k \in \mathcal{K}\), where
\begin{equation} \label{b^out2}
	\tilde{b}_{k,m}^{**} = \begin{cases}
		0, \ \text{if} \ \mu_k > 0 \ \text{and} \ \bar{\Gamma}_{m,2} \le \Gamma - \frac{\|\boldsymbol{w}_m\|^2 \sigma_z^2}{K^2} < \frac{1}{K}, \\
		\frac{|{\boldsymbol{w}_m}^H \hat{\boldsymbol{h}}_{k,m}|}{|{\boldsymbol{w}_m}^H \hat{\boldsymbol{h}}_{k,m}|^2 + \|\boldsymbol{w}_m\|^2 \sigma_{e,k}^2 + \frac{\mu_k}{\lambda_m^{**}}}, \ \text{otherwise}.
	\end{cases}
\end{equation}
Here, \(\lambda_m^{**}\) is obtained by applying the bisection search based on the following equality. 
\begin{equation} \label{lam2}
	\begin{aligned}
		\sum_{k \in \mathcal{K}} \frac{(\|\boldsymbol{w}_m\|^2 \sigma_{e,k}^2 + \frac{\mu_k}{\lambda_m^{**}})^2 + |{\boldsymbol{w}_m}^H \hat{\boldsymbol{h}}_{k,m}|^2 \|\boldsymbol{w}_m\|^2 \sigma_{e,k}^2}{(|{\boldsymbol{w}_m}^H \hat{\boldsymbol{h}}_{k,m}|^2 + \|\boldsymbol{w}_m\|^2 \sigma_{e,k}^2 + \frac{\mu_k}{\lambda_m^{**}})^2} \\ 
		+ \|\boldsymbol{w}_m\|^2 \sigma_z^2 = K^2 \Gamma.
	\end{aligned}			 
\end{equation}

\textit{Proof}: See Appendix C.
\hfill \(\square\)

Accordingly, problem (P4.\(m\)) can be optimally solved via the following proposition, which can be proved similarly as Proposition 3 for problem (P3.\(m\)) in the SISO case.

\textit{Proposition 6}:
Let \(\{\tilde{b}_{k,m}^{\star\star}\}\) denote the optimal solution to (P4.\(m\)). For each subcarrier \(m\) with \(\sum_{k \in \mathcal{K}} \mu_k (\tilde{b}_{k,m}^{**})^2 > 1\), the optimal solution to (P4.\(m\)) is \(\{\tilde{b}_{k,m}^{\star\star} = 0\}\); for each subcarrier \(m\) with \(\sum_{k \in \mathcal{K}} \mu_k (\tilde{b}_{k,m}^{**})^2 \le 1\), the optimal solution to (P4.\(m\)) is \(\{\tilde{b}_{k,m}^{\star\star} = \tilde{b}_{k,m}^{**}\}\).


Next, we utilize subgradient-based methods such as the ellipsoid method \cite{boyd2004convex} to solve the dual problem (D2.3). Utilizing the fact that the subgradient of objective function \(g_3\big(\{\mu_k\}\big)\) at \(\mu_k\) is \(\sum_{m \in \mathcal{M}} (\tilde{b}_{k,m}^{\star\star})^2 - P_k, \forall k \in \mathcal{K}\), the ellipsoid method can efficiently obtain the optimal dual solution to (D2.3) as \(\{\mu_k^{\mathrm{IV}}\}\).

Then, we present the optimal solution to the primal problem (P2.3). By replacing \(\{\mu_k\}\) as the optimal dual variables \(\{\mu_k^{\mathrm{IV}}\}\), the optimal solution \(\{\tilde{b}_{k,m}^{\star\star}\}\) to problem \eqref{dual3} in Proposition 6 becomes the optimal primal solution to (P2.3) with given \(\{\boldsymbol{w}_m\}\), denoted by \(\{\tilde{b}_{k,m}^{\mathrm{IV}}\}\).

\textit{Remark 4}: It is worth noting that for the general SIMO case, the optimized transmit amplitude solution (or power control policy) in \eqref{b^out2} follows an on-off regularized channel inversion structure, similar to that in Remark 2 for the special SISO case.

\subsubsection{Complete Algorithm for Solving Problem (P2.2)}

Finally, the alternating-optimization-based algorithm for solving (P2.2) is implemented in an iterative manner. In each iteration, we first update the transmit coefficients as \(\{\tilde{b}_{k,m}^{\mathrm{IV}}\}\) in \eqref{b^out2} under given \(\{\boldsymbol{w}_m\}\), and then update the receive beamforming vector as \(\{\boldsymbol{w}_m^{\mathrm{IV}}\}\) based on \eqref{w^out} under given \(\{\tilde{b}_{k,m}\}\).


It is interesting to discuss the case when  each WD has asymptotically high transmit power (i.e., \( P_k \to \infty, \forall k\in\mathcal K\)). According to Proposition 4, if \(\Gamma \ge \bar{\Gamma}_{m,1}, \forall m \in \mathcal{M}\), then \(\mathrm{MSE}_{\mathrm{out}} = 0\); otherwise, if \(\Gamma < \bar{\Gamma}_{m,1}\) for any subcarrier \(m \in \mathcal{M}\), then the computation outage becomes inevitable. 
Furthermore, we consider the case when the AP is equipped with a massive number of receive antennas (i.e., \(N_r \to \infty\)) and the channel vectors are i.i.d.. In this case, according to Proposition 5, we have \(\mathrm{MSE}_m \to 0, \forall m \in \mathcal{M}\). This implies that no matter how small the MSE threshold \(\Gamma\) is, we can achieve a zero outage probability by exploiting the receive beamforming with massive antennas, which highlights the benefit of massive antennas for AirComp again. 

\section{Numerical Results}

This section evaluates the AirComp performance of our proposed designs. We consider the following three benchmarks for performance comparison.

\begin{itemize}
	\item Benchmark ignoring CSI errors:
	The WDs and AP optimize the transceiver design via solving problems (P1) and (P2) by ignoring the channel estimation errors.
	
	\item Equal power allocation:
	Each WD \(k \in \mathcal{K}\) transmits with the full power and aligned phase, where the power is equally allocated among its \(M\) subcarriers, i.e., \(b_{k,m} = \sqrt{\frac{P_k}{M}} \frac{\hat{\boldsymbol{h}}_{k,m}^H \boldsymbol{w}_m}{|\boldsymbol{w}_m^H \hat{\boldsymbol{h}}_{k,m}|}\).
	
	\item Channel inversion power control:
	Each WD \(k \in \mathcal{K}\) sets the transmit coefficients for its \(M\) subcarriers based on the channel inversion principle, i.e., \(b_{k,m} = \sqrt{\frac{P_k}{M}} \frac{\min_{i \in \mathcal{K}} \|\hat{\boldsymbol{h}}_{i,m}\|}{\sqrt{\|\hat{\boldsymbol{h}}_{k,m}\|^2 + \sigma_{e,k}^2}} \frac{\hat{\boldsymbol{h}}_{k,m}^H \boldsymbol{w}_m}{|\boldsymbol{w}_m^H \hat{\boldsymbol{h}}_{k,m}|}\). 
\end{itemize}
Notice that in the last two schemes, the receive beamforming vectors \(\{\boldsymbol{w}_m\}\) at the AP are designed similarly as that in Sections III and IV for SISO and SIMO cases, respectively.

In the simulation, we set the channel vectors \(\boldsymbol{h}_{k,m}\)'s as independent CSCG random vectors with zero mean and covariance \(\sigma_{h,k}^2 \boldsymbol{I}\), where \(\sigma_{h,k}^2\)'s are randomly generated to capture the differences of large-scale Rayleigh fading at different WDs. We also set \(P_k = P\) and \(\sigma_{e,k}^2 = \sigma_{e}^2\), \(\forall k \in \mathcal{K}\).

\begin{figure}[tb]
	\centering 
	\subfloat[\footnotesize Best-effort computation \newline scenario.] {\includegraphics[width=0.5\columnwidth]{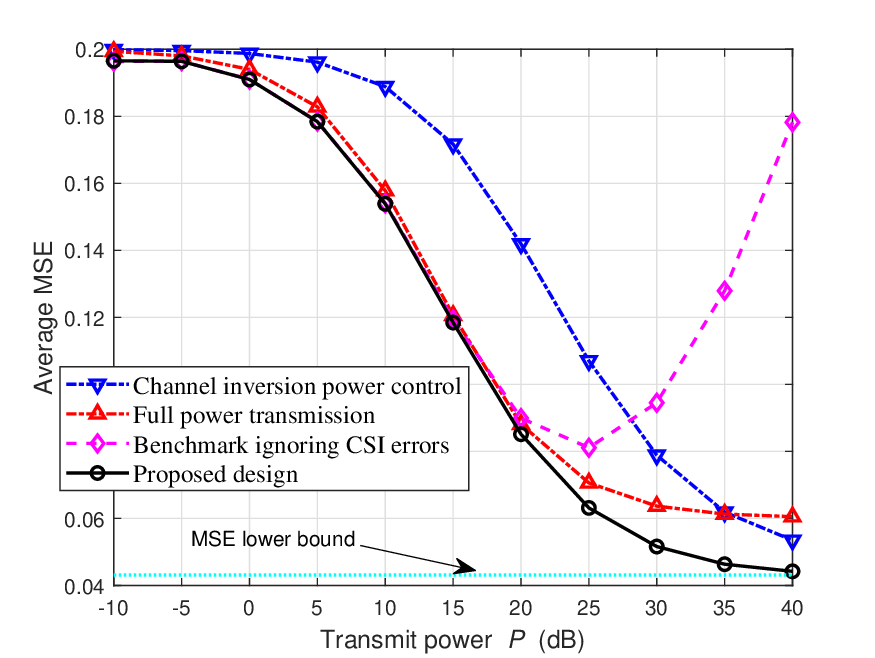}} 
	\subfloat[\footnotesize Error-constrained computation \newline scenario.] {\includegraphics[width=0.5\columnwidth]{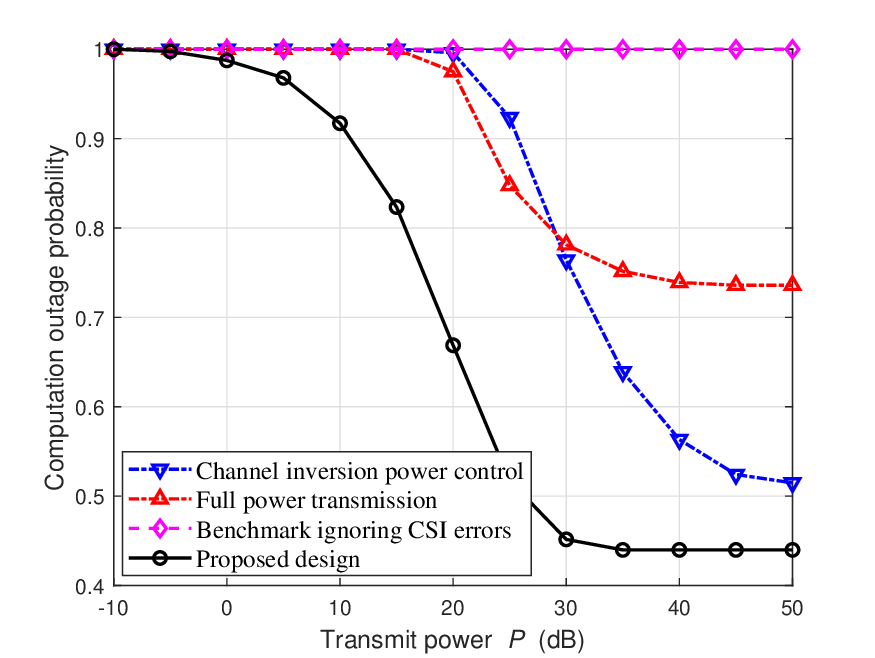}} 
	\caption{The average MSE and computation outage probability versus transmit power \(P\) for the SISO case with \(K = 5\), \(M = 128\), \(\sigma_e^2 = 0.2\), and \(\Gamma = 0.05\).} 
\end{figure}
Figs. 2(a) and 2(b) show the average MSE and computation outage probability versus the transmit power \(P\) at each WD in the SISO case for the best-effort computation and error-constrained computation scenarios, respectively, where \(K = 5\), \(M = 128\), \(\sigma_e^2 = 0.2\), and \(\Gamma = 0.05\). It is observed that the proposed design outperforms the other benchmarks across the entire transmit power regime. In Fig. 2(a), we show the MSE lower bound for comparison, which corresponds to \(\frac{1}{M} \sum_{m \in \mathcal{M}} \Gamma_{m,1}\). When \(P\) becomes large, the computation MSE achieved by the proposed design is observed to approach the lower bound, as indicated in Proposition 2. In the low transmit power regime, the equal power allocation is observed to perform close to the proposed design, as it can efficiently suppress the noise-induced error that dominates the MSE in this case. In the high transmit power regime, the channel inversion power control is observed to perform close to the proposed design, due to the efficient signal magnitude alignment. For the best-effort computation scenario in Fig. 2(a), the benchmark ignoring CSI errors exhibits the poorest performance and leads to an increased MSE when \(P\) becomes large. This is due to the fact that the CSI errors are amplified by the high transmit power, hence deteriorating the MSE performance. For the error-constrained computation scenario in Fig. 2(b), the benchmark ignoring CSI errors leads to a computation outage probability of one in the whole transmit power regime. This is because this scheme only considers the signal misalignment error and noise-induced error terms in \eqref{mse} when minimizing the MSE, and the actually achieved MSE will exceed the threshold when the CSI-related error term is further taken into account. 

\begin{figure}[tb]
	\centering 
	\subfloat[\footnotesize Best-effort computation \newline scenario.] {\includegraphics[width=0.5\columnwidth]{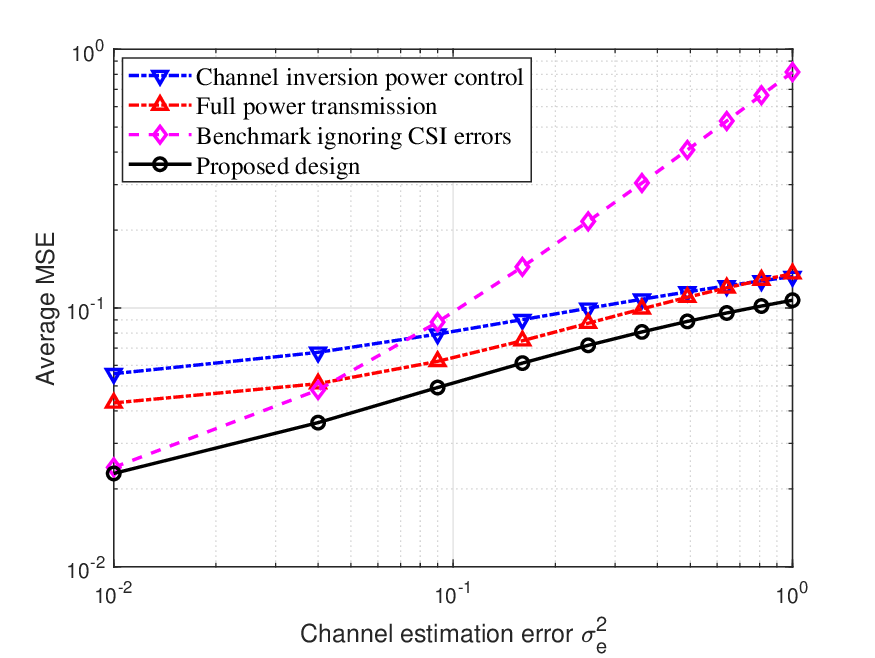}} 
	\subfloat[\footnotesize Error-constrained computation \newline scenario.] {\includegraphics[width=0.5\columnwidth]{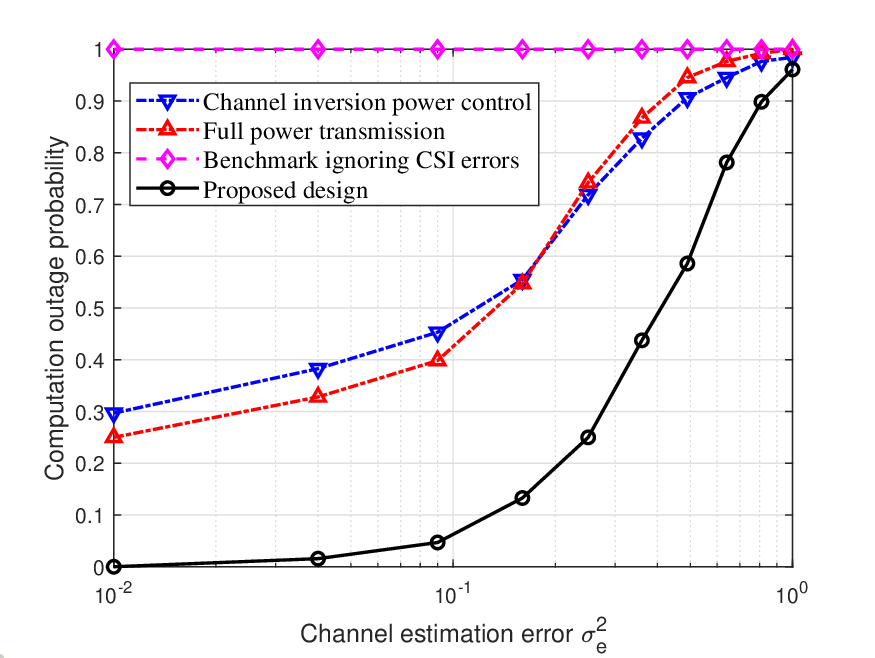}} 
	\caption{The average MSE and computation outage probability versus channel estimation error \(\sigma_e^2\) for the SISO case with \(P = 30\) dB, \(K = 5\), \(M = 128\), and \(\Gamma = 0.05\).} 
\end{figure}
Figs. 3(a) and 3(b) show the average MSE and computation outage probability versus the variance of CSI error \(\sigma_e^2\) in the SISO case for the best-effort computation and error-constrained computation scenarios, respectively, where \(P = 30\) dB, \(K = 5\), \(M = 128\), and \(\Gamma = 0.05\). It is observed that the proposed design outperforms the other benchmarks across the entire regime of \(\sigma_{e}^2\). The performance gap between the proposed design and equal power allocation/channel inversion power control diminishes as \(\sigma_e^2\) increases. This is because with larger CSI errors, the impact of power control becomes less significant, and these schemes are more influenced by the CSI-related error. For the best-effort computation scenario in Fig. 3(a), the benchmark ignoring CSI errors is observed to perform far worse than the other benchmarks when \(\sigma_{e}^2\) becomes large. For the error-constrained computation scenario in Fig. 3(b), the benchmark ignoring CSI errors leads to the computation outage probability of one in the whole transmit power regime, which can be similarly explained as for Fig. 2(b).

\begin{figure}[tb]
	\centering 
	\subfloat[\footnotesize Best-effort computation \newline scenario.] {\includegraphics[width=0.5\columnwidth]{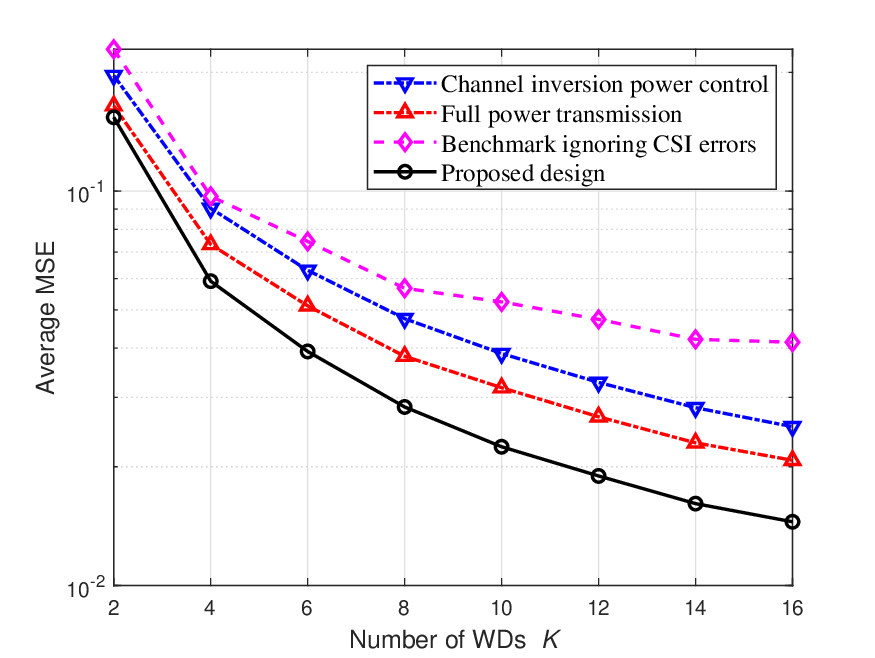}} 
	\subfloat[\footnotesize Error-constrained computation \newline scenario.] {\includegraphics[width=0.5\columnwidth]{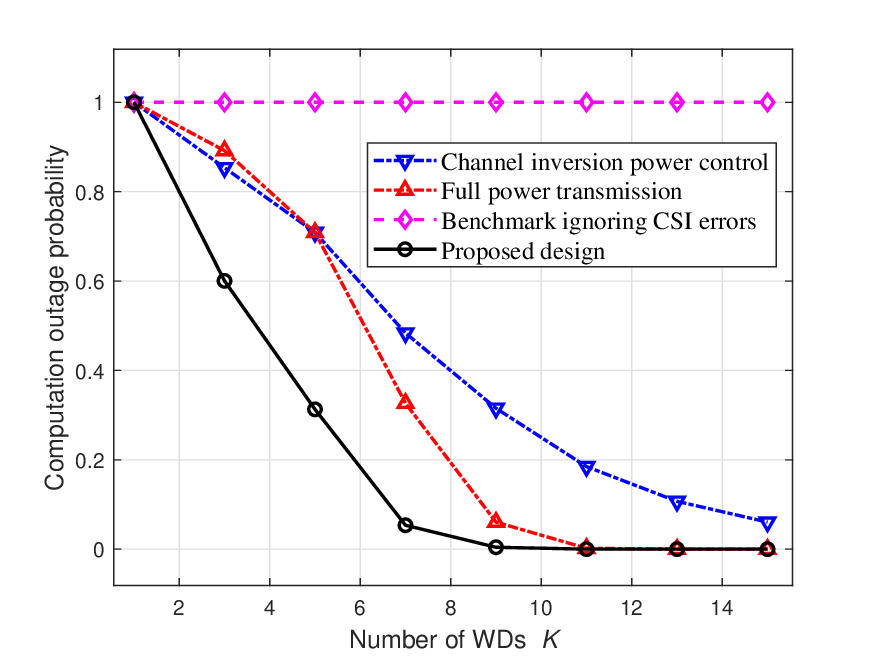}} 
	\caption{The average MSE and computation outage probability versus the number of WDs \(K\) for the SISO case with \(P = 30\) dB, \(M = 128\), \(\sigma_e^2 = 0.2\), and \(\Gamma = 0.05\).} 
\end{figure}
Figs. 4(a) and 4(b) show the average MSE and computation outage probability versus the number of WDs \(K\) in the SISO case for the best-effort computation and error-constrained computation scenarios, respectively, where \(P = 30\) dB, \(M = 128\), \(\sigma_e^2 = 0.2\), and \(\Gamma = 0.05\). It is observed that the computation MSEs achieved by all four schemes decrease as \(K\) increases, due to the fact that the AP can aggregate more data for averaging. For the best-effort computation scenario in Fig. 4(a), the performance gap between the proposed design and the three benchmarks is observed to become more significant as \(K\) increases, indicating the effectiveness of the proposed scheme in managing CSI errors.


\begin{figure}[tb]
	\centering 
	\subfloat[\footnotesize Best-effort computation \newline scenario.] {\includegraphics[width=0.5\columnwidth]{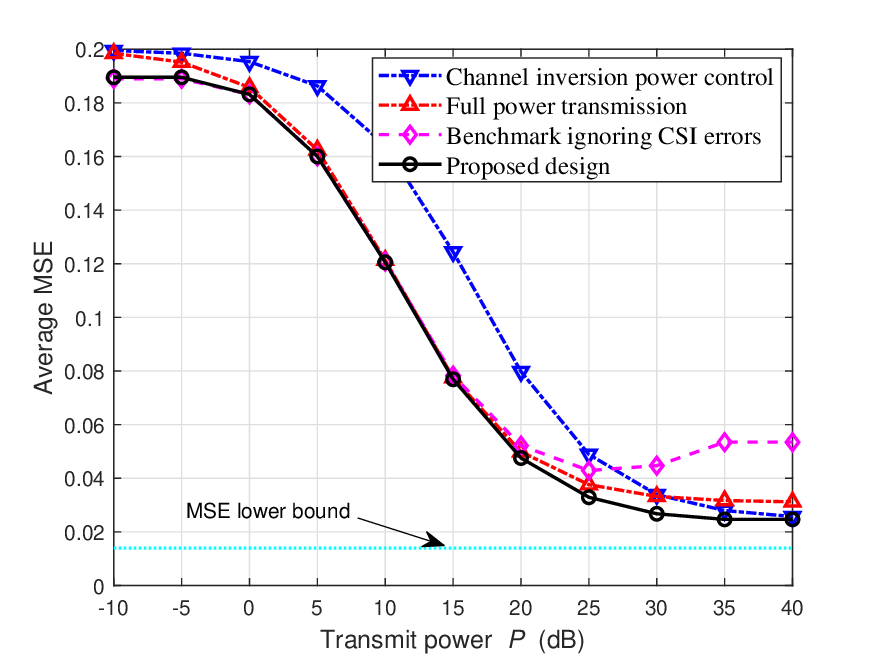}} 
	\subfloat[\footnotesize Error-constrained computation \newline scenario.] {\includegraphics[width=0.5\columnwidth]{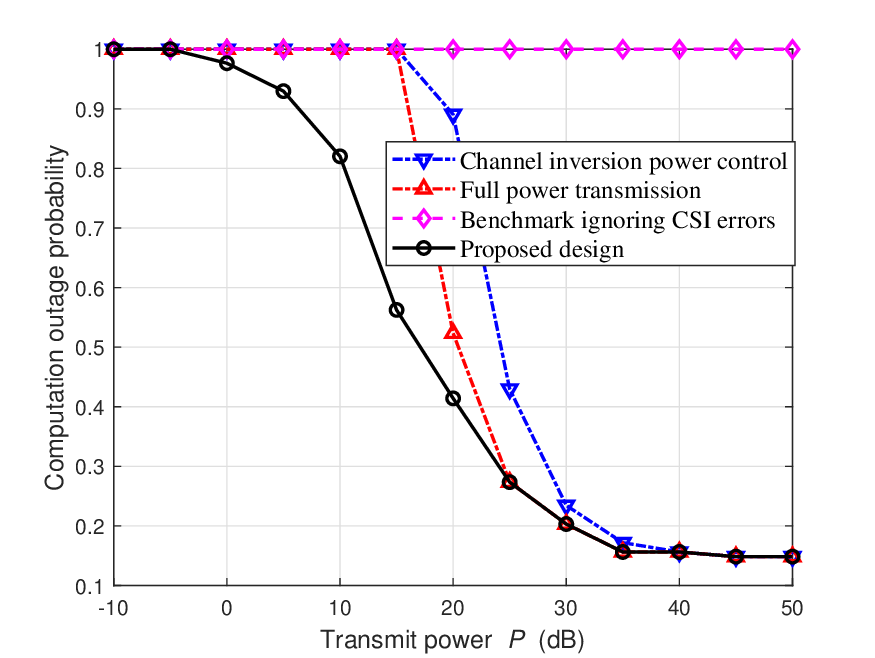}} 
	\caption{The average MSE and computation outage probability versus transmit power \(P\) for the SIMO case with \(N_r = 4\), \(K = 5\), \(M = 128\), \(\sigma_e^2 = 0.2\), and \(\Gamma = 0.05\).} 
\end{figure}
Figs. 5(a) and 5(b) show the average MSE and computation outage probability versus the transmit power \(P\) at each WD in the SIMO case with \(N_r = 4\) for the best-effort computation and error-constrained computation scenarios, respectively, where \(K = 5\), \(M = 128\), \(\sigma_e^2 = 0.2\), and \(\Gamma = 0.05\). Similar to the observations in the SISO case, it is observed that in the low power regime, the equal power allocation performs close to the proposed design, while in the high power regime, the channel inversion power control performs close to the proposed design. The benchmark ignoring CSI errors is observed to perform the worst and even lead to an increased MSE when \(P\) becomes large. In Fig. 5(a), we show the MSE lower bound, which corresponds to \(\frac{1}{M} \sum_{m \in \mathcal{M}} \bar{\Gamma}_{m,1}\) in Proposition 4. When \(P\) becomes large, it is observed that there exists a gap between the computation MSE achieved by the proposed design versus the lower bound, which is due to the sub-optimality of the proposed design.

\begin{figure}[tb]
	\centering 
	\subfloat[\footnotesize Best-effort computation \newline scenario.] {\includegraphics[width=0.5\columnwidth]{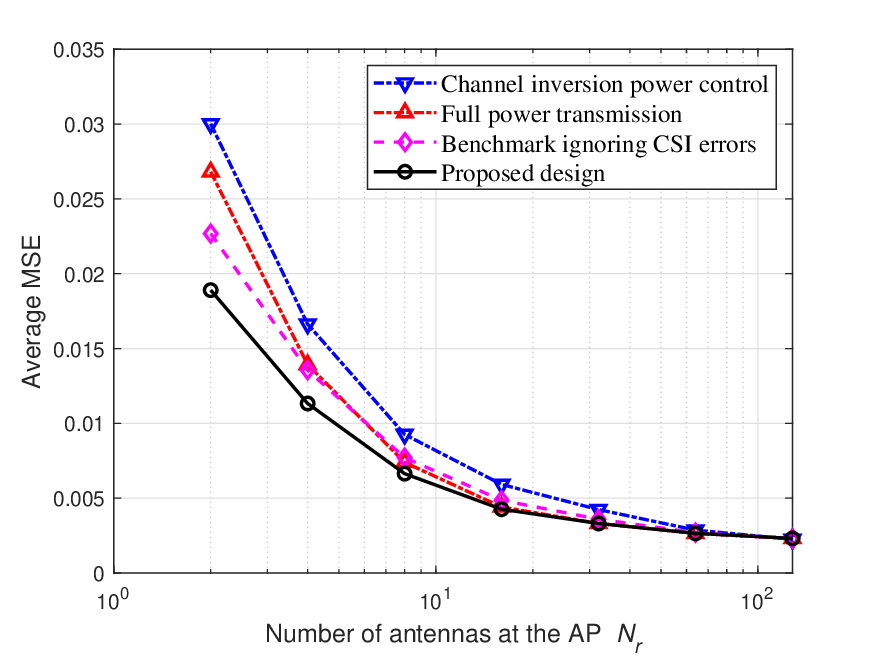}} 
	\subfloat[\footnotesize Error-constrained computation \newline scenario.] {\includegraphics[width=0.5\columnwidth]{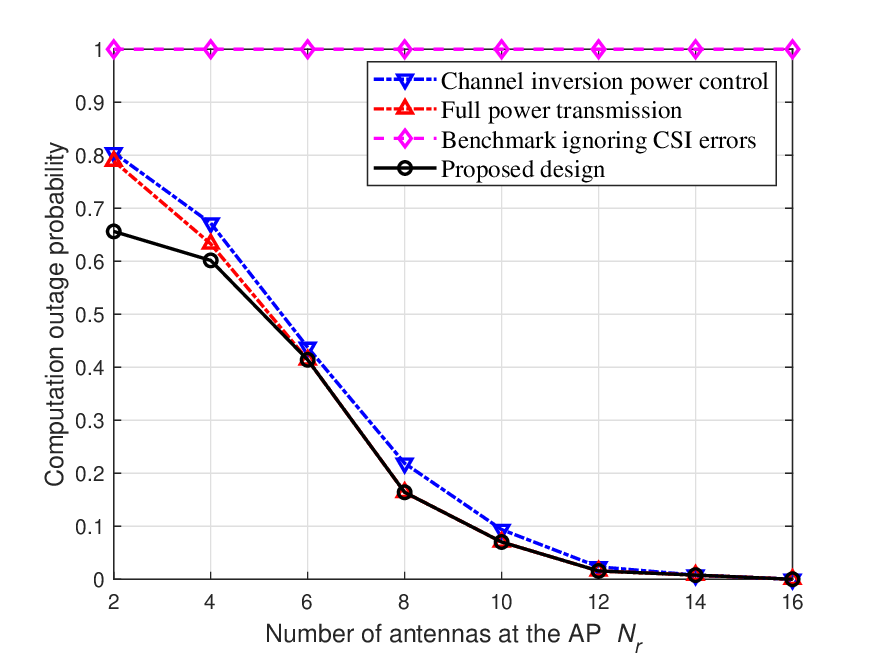}} 
	\caption{The average MSE and computation outage probability versus the number of antennas at the AP \(N_r\) for the SIMO case with \(P = 30\) dB, \(K = 5\), \(M = 128\), \(\sigma_e^2 = 0.2\), and \(\Gamma = 0.05\).} 
\end{figure}
Figs. 6(a) and 6(b) show the average MSE and computation outage probability versus the number of receive antennas \(N_r\) at the AP in the SIMO case for the best-effort computation and error-constrained computation scenarios, respectively, where \(P = 30\) dB, \(K = 5\), \(M = 128\), \(\sigma_e^2 = 0.2\), and \(\Gamma = 0.05\).
It is observed that as \(N_r\) increases, the performances achieved by the equal power allocation and the channel inversion power control approach that by the proposed design. This is due to the fact that the receive beamforming becomes more critical for data aggregation in this case, and thus the gain provided by transmit power control becomes marginal. 

\section{Conclusion}

This paper considered the joint transceiver design to minimize the computation MSE for an uncoded OFDM AirComp system with imperfect CSI. We considered two scenarios with best-effort and error-constrained computation tasks, with the objectives of minimizing the average MSE and the computation outage probability over subcarriers, respectively. For the SISO case, we derived the optimal (on-off) regularized channel inversion power control solutions to the two problems. For the SIMO case, we proposed alternating-optimization-based algorithms to find high-quality solutions. In addition, we derived interesting analytic results on the computation MSE in the regimes with asymptotically high transmit powers or an asymptotically large number of receive antennas. Remarkable MSE performance gains were observed by our proposed designs, in comparison with benchmark schemes, which showed the importance of jointly optimizing the transmit power control and the receive strategy to combat against channel estimation errors for reliable AirComp.

\appendix

\subsection{Proof of Proposition 1}

In the following, we first obtain the dual function \(g_1\big(\{\mu_k\}\big)\) with given dual variables, and then search over \(\{\mu_k \ge 0\}\) to minimize \(g_1\big(\{\mu_k\}\big)\).

First, we find the dual function \(g_1\big(\{\mu_k\}\big)\) in \eqref{dual1} under given \(\{\mu_k \ge 0\}\). By dropping the constant term \(- \sum_{k \in \mathcal{K}} \mu_k P_k\), the problem in \eqref{dual1} can be decomposed into the following \(M\) subproblems, each for one subcarrier \(m \in \mathcal{M}\).
\begin{equation*}
	\begin{aligned}
		(\text{P5}.m): \min_{\{\tilde{b}_{k,m}\}, w_m} \ &\sum_{k \in \mathcal{K}} \big((w_m |\hat{h}_{k,m}| \tilde{b}_{k,m} - 1)^2 \\
		&+ w_m^2 \sigma_{e,k}^2 \tilde{b}_{k,m}^2\big) + w_m^2 \sigma_z^2 + \sum_{k \in \mathcal{K}} \mu_k \tilde{b}_{k,m}^2.
	\end{aligned}
\end{equation*}

To solve problem (P5.\(m\)), we first optimize \(\{\tilde{b}_{k,m}\}\) under given \(w_m\). In this case, (P5.\(m\)) can be further decomposed into the following \(K\) subproblems, each for one WD \(k \in \mathcal{K}\).
\begin{equation*}
	(\text{P5}.m.k): \min_{\tilde{b}_{k,m}} \ (w_m |\hat{h}_{k,m}| \tilde{b}_{k,m} - 1)^2 + w_m^2 \sigma_{e,k}^2 \tilde{b}_{k,m}^2 + \mu_k \tilde{b}_{k,m}^2.
\end{equation*}
By checking the first derivative of the objective function, the optimal solution of \(\{\tilde{b}_{k,m}\}\) to problem (P5.\(m\).\(k\)) is obtained as
\begin{equation} \label{b^*}
	\tilde{b}_{k,m}^{***} = \frac{w_m |\hat{h}_{k,m}|}{w_m^2 |\hat{h}_{k,m}|^2 + w_m^2 \sigma_{e,k}^2 + \mu_k}.
\end{equation}
Then, we optimize \(w_m\). By substituting \(\{\tilde{b}_{k,m}^{***}\}\) back to (P5.\(m\)), the optimization of \(w_m\) is expressed as
\begin{equation} \label{5.2}
	\min_{w_m} \ \sum_{k \in \mathcal{K}} \frac{w_m^2 \sigma_{e,k}^2 + \mu_k}{w_m^2 |\hat{h}_{k,m}|^2 + w_m^2 \sigma_{e,k}^2 + \mu_k} \\ + w_m^2 \sigma_z^2 \triangleq G(w_m).
\end{equation}
By setting the gradient \(\frac{\partial G(w_m)}{\partial w_m^2}\) to be zero, the optimal solution of \(w_m\) to problem \eqref{5.2} should satisfy
\begin{equation} \label{w^*}
	\sum_{k \in \mathcal{K}} \frac{|\hat{h}_{k,m}|^2 \mu_k}{\big((w_m^{***})^2 (|\hat{h}_{k,m}|^2 + \sigma_{e,k}^2) + \mu_k\big)^2} = \sigma_z^2,
\end{equation}
where the left-hand side (L.H.S.) is monotonically decreasing w.r.t. \(w_m^{***}\). Therefore, we can find \(w_m^{***}\) by using a bisection search based on \eqref{w^*} efficiently.
By substituting \(w_m^{***}\) into \eqref{b^*}, we get \(\{\tilde{b}_{k,m}^{***}\}\) w.r.t. \(\{\mu_k\}\). Therefore, subproblems (P5.\(m\)) are finally solved under given \(\{\mu_k\}\), and the dual function \(g_1\big(\{\mu_k\}\big)\) in \eqref{dual1} is accordingly obtained.

Next, we solve the dual problem (D1.1) by applying subgradient-based methods such as the ellipsoid method \cite{boyd2004convex}. The subgradient of objective function \(g_1\big(\{\mu_k\}\big)\) at \(\mu_k\) is \(\sum_{m \in \mathcal{M}} (\tilde{b}_{k,m}^{***})^2 - P_k, \forall k \in \mathcal{K}\). Using this subgradient, the ellipsoid method can obtain the optimal dual solution to (D1.1) as \(\{\mu_k^{\mathrm{I}}\}\).

Now, we present the optimal solution to the primal problem (P1.1). With the optimal dual variables \(\{\mu_k^{\mathrm{I}}\}\) at hand, the optimal solution \(\{\tilde{b}_{k,m}^{***}\}\) and \(\{w_m^{***}\}\) in \eqref{b^*} and \eqref{w^*} to problem \eqref{dual1} can be directly used for constructing the optimal primal solution \(\{\tilde{b}_{k,m}^{\mathrm{I}}\}\) and \(\{w_m^{\mathrm{I}}\}\) to (P1.1).

\subsection{Proof of Lemma 1}


To solve problem (P3.1.\(m\)), for each WD \(k\) with \(\mu_k = 0\), according to Proposition 2, we set \(\{\tilde{b}_{k,m}^* = \frac{|\hat{h}_{k,m}|}{w_m |\hat{h}_{k,m}|^2 + w_m \sigma_{e,k}^2}\}_{k \in \mathcal{K}, \mu_k = 0}\) to minimize the computation MSE without affecting the objective. Therefore, solving (P3.1.\(m\)) remains to optimize \(\tilde{b}_{k,m}\)'s with \(\mu_k > 0\) by the following problem.
\begin{equation*}
	\begin{aligned}
		(\text{P3.2}.m): & \min_{\{\tilde{b}_{k,m}\}_{k \in \mathcal{K}, \mu_k > 0}, w_m} \ \sum_{k \in \mathcal{K}, \mu_k > 0} \mu_k \tilde{b}_{k,m}^2 \\
		\mathrm{s.t.} &\ \sum_{k \in \mathcal{K}, \mu_k > 0} \big((w_m |\hat{h}_{k,m}| \tilde{b}_{k,m} - 1)^2 + w_m^2 \sigma_{e,k}^2 \tilde{b}_{k,m}^2\big) \\
		&\ + \sum_{k \in \mathcal{K}, \mu_k = 0} \frac{\sigma_{e,k}^2}{|\hat{h}_{k,m}|^2 + \sigma_{e,k}^2} + w_m^2 \sigma_z^2 \le K^2 \Gamma.
	\end{aligned}
\end{equation*}

For problem (P3.2.\(m\)), we first consider the case with \(\Gamma_{m,2} < \Gamma < \frac{1}{K}\), for which we can simply set \(\{\tilde{b}_{k,m}^* = 0\}_{k \in \mathcal{K}, \mu_k > 0}\) and \(w_m^* = \frac{K}{\sigma_z} \sqrt{\Gamma-\Gamma_{m,2}}\) to make the MSE constraint hold. 

\textit{Lemma 3}: For the case with \(\Gamma_{m,2} < \Gamma < \frac{1}{K}\), if problem (P3.2.\(m\)) is feasible, then with finite transmit power, it cannot happen that \(\Gamma_{m,2} \to \Gamma\). 

\textit{Proof}:
If so, then according to the discussion above, the MSE constraint in (P3.2.\(m\)) in this case becomes \(K^2 \Gamma_{m,2} + w_m^2 \sigma_z^2 \le K^2 \Gamma\), which implies \(w_m^* \to 0\) and \(\{\tilde{b}_{k,m}^* \to \infty\}_{k \in \mathcal{K}, \mu_k = 0}\).
\hfill \(\square\)

Next, we consider the case with \(\Gamma_{m,1} < \Gamma \le \Gamma_{m,2}\). Let \(\{\lambda_m \ge 0\}\) denote the dual variable associated with the MSE constraint in (P3.2.\(m\)). The Lagrangian of (P3.2.\(m\)) is
\begin{equation} \label{ll}
	\begin{aligned}
		\mathcal{L}_4 & \big(\{\tilde{b}_{k,m}\}_{k \in \mathcal{K}, \mu_k > 0}, w_m, \lambda_m\big) = \sum_{k \in \mathcal{K}, \mu_k > 0} \mu_k \tilde{b}_{k,m}^2 \\
		&+ \lambda_m \Big(\sum_{k \in \mathcal{K}, \mu_k > 0} \big((w_m |\hat{h}_{k,m}| \tilde{b}_{k,m} - 1)^2 + w_m^2 \sigma_{e,k}^2 \tilde{b}_{k,m}^2\big)\\
		&+ \sum_{k \in \mathcal{K}, \mu_k = 0} \frac{\sigma_{e,k}^2}{|\hat{h}_{k,m}|^2 + \sigma_{e,k}^2} + w_m^2 \sigma_z^2 - K^2 \Gamma \Big),
	\end{aligned}	
\end{equation}
whose first derivative for \(\{\tilde{b}_{k,m}\}\) equals to zero when
\begin{equation} \label{bb}
	\tilde{b}_{k,m} = \frac{\lambda_m w_m |\hat{h}_{k,m}|}{\lambda_m w_m^2 |\hat{h}_{k,m}|^2 + \lambda_m w_m^2 \sigma_{e,k}^2 + \mu_k}, \forall k \in \mathcal{K}, \mu_k > 0.
\end{equation}

\textit{Lemma 4}: For the case with \(\Gamma_{m,1} < \Gamma \le \Gamma_{m,2}\), if (P3.2.\(m\)) is feasible, then with finite transmit power, there is at least one WD \(k\) satisfying that \(\mu_k > 0\), and it follows that \(\lambda_m > 0\) and \(w_m > 0\). 

\textit{Proof}:
If \(\mu_k = 0, \forall k \in \mathcal{K}\), then we have \(\Gamma_{m,1} = \Gamma_{m,2}\). If either \(\lambda_m = 0\) or \(w_m = 0\), then according to \eqref{bb}, we have \(\{\tilde{b}_{k,m}^* = 0\}_{k \in \mathcal{K}, \mu_k > 0}\), thus the MSE constraint becomes \(K^2 \Gamma_{m,2} + w_m^2 \sigma_z^2 \le K^2 \Gamma\). Both are in contradiction with the fact \(\Gamma_{m,1} < \Gamma \le \Gamma_{m,2}\).
\hfill \(\square\)

Substituting \eqref{bb} into \eqref{ll}, when the gradient \(\frac{\partial \mathcal{L}_4(\{\tilde{b}_{k,m}\}_{k \in \mathcal{K}, \mu_k > 0}, w_m, \lambda_m)}{\partial w_m^2}\) equals to zero, it follows that \(\lambda_m \big(\sum_{k \in \mathcal{K}, \mu_k > 0} \frac{\lambda_m |\hat{h}_{k,m}|^2 \mu_k}{(\lambda_m w_m^2 |\hat{h}_{k,m}|^2 + \lambda_m w_m^2 \sigma_{e,k}^2 + \mu_k)^2} - \sigma_z^2\big) = 0\),
which is equivalent to the equality as follows as \(\lambda_m > 0\) and \(w_m > 0\) hold.
\begin{equation} \label{ww}
	\sum_{k \in \mathcal{K}, \mu_k > 0} \frac{v_m |\hat{h}_{k,m}|^2 \mu_k}{(v_m |\hat{h}_{k,m}|^2 + v_m \sigma_{e,k}^2 + \mu_k)^2} = w_m^2 \sigma_z^2,
\end{equation}
where \(v_m \triangleq \lambda_m w_m^2 > 0\). Substituting \eqref{ww} back, the MSE constraint in (P3.2.\(m\)) w.r.t. \(\lambda_m\) becomes
\begin{equation} \label{llam}
	\begin{aligned}
		\sum_{k \in \mathcal{K}} \frac{v_m \sigma_{e,k}^2 + \mu_k}{v_m |\hat{h}_{k,m}|^2 + v_m\sigma_{e,k}^2 + \mu_k} \le K^2 \Gamma,
	\end{aligned}
\end{equation}
whose L.H.S. is monotonously decreasing from \(K^2 \Gamma_{m,2}\) to \(K^2 \Gamma_{m,1}\) as \(v_m\) increases from \(0\) to infinity. Since \(\lambda_m > 0\), \eqref{llam} is tight at the optimality. Therefore, we can determine \(v_m^*\) based on the equality of \eqref{llam} by applying the bisection search. 

\textit{Lemma 5}: For the case with \(\Gamma_{m,1} < \Gamma \le \Gamma_{m,2}\), if problem (P3.2.\(m\)) is feasible, then with finite transmit power, it cannot happen that \(\Gamma_{m,2} = \Gamma\). 

\textit{Proof}:
If so, then according to the previous discussion, we have \(v_m^* = 0\), which is in contradiction with Lemma 4.
\hfill \(\square\)

Finally, by substituting \(v_m^*\) into \eqref{ww}, we get the optimal solution of \(w_m\) to (P3.2.\(m\)), and thus (P3.1.\(m\)) as \(w_m^*\), and accordingly obtain the optimal dual variable \(\lambda_m\) as \(\lambda_m^* = v_m^*/(w_m^*)^2\).
Substituting \(\lambda_m^*\) and \(w_m^*\) back to \eqref{bb}, we obtain the optimal solution of \(\{\tilde{b}_{k,m}\}_{k \in \mathcal{K}, \mu_k > 0}\) as \(\{\tilde{b}_{k,m}^*\}_{k \in \mathcal{K}, \mu_k > 0}\). This thus completes the proof.

\subsection{Proof of Lemma 2}

Similar to problem (P3.1.\(m\)) for the SISO case, we reexpress problem (P4.1.\(m\)) as follows by setting \(\{\tilde{b}_{k,m}^{**} = \frac{|{\boldsymbol{w}_m}^H \hat{\boldsymbol{h}}_{k,m}|}{|{\boldsymbol{w}_m}^H \hat{\boldsymbol{h}}_{k,m}|^2 + \|\boldsymbol{w}_m\|^2 \sigma_{e,k}^2}\}_{k \in \mathcal{K}, \mu_k = 0}\),
\begin{equation*}
	\begin{aligned}
		(\text{P4.2}.&m): \min_{\{\tilde{b}_{k,m}\}_{k \in \mathcal{K}, \mu_k > 0}} \ \sum_{k \in \mathcal{K}, \mu_k > 0} \mu_k \tilde{b}_{k,m}^2 \\
		\mathrm{s.t.} &\ \sum_{k \in \mathcal{K}, \mu_k > 0} \big((|\boldsymbol{w}_m^H \hat{\boldsymbol{h}}_{k,m}| \tilde{b}_{k,m} - 1)^2 + \|\boldsymbol{w}_m\|^2 \sigma_{e,k}^2 \tilde{b}_{k,m}^2\big) \\
		&\ + \sum_{k \in \mathcal{K}, \mu_k = 0} \frac{\|\boldsymbol{w}_m\|^2 \sigma_{e,k}^2}{|\boldsymbol{w}_m^H \hat{\boldsymbol{h}}_{k,m}|^2 + \|\boldsymbol{w}_m\|^2 \sigma_{e,k}^2} \\
		&\ + \|\boldsymbol{w}_m\|^2 \sigma_z^2 \le K^2 \Gamma,
	\end{aligned}
\end{equation*}
and when \(\Gamma \ge \bar{\Gamma}_{m,2} + \frac{\|\boldsymbol{w}_m\|^2 \sigma_z^2}{K^2}\), setting \(\{\tilde{b}_{k,m}^{**} = 0\}_{k \in \mathcal{K}, \mu_k > 0}\) can make the MSE constraint hold.

When \(\bar{\Gamma}_{m,1} < \Gamma - \frac{\|\boldsymbol{w}_m\|^2 \sigma_z^2}{K^2} < \bar{\Gamma}_{m,2}\), let \(\{\lambda_m \ge 0\}\) denote the dual variable associated with the MSE threshold constraint in problem (P4.2.\(m\)). The Lagrangian of (P4.2.\(m\)) is given by \(\mathcal{L}_5(\{\tilde{b}_{k,m}\}, \lambda_m) = \sum_{k \in \mathcal{K}} \mu_k \tilde{b}_{k,m}^2 + \lambda_m \Big(\sum_{k \in \mathcal{K}, \mu_k > 0} \big((|\boldsymbol{w}_m^H \hat{\boldsymbol{h}}_{k,m}| \tilde{b}_{k,m} - 1)^2 + \|\boldsymbol{w}_m\|^2 \sigma_{e,k}^2 \tilde{b}_{k,m}^2\big) + \sum_{k \in \mathcal{K}, \mu_k = 0} \frac{\|\boldsymbol{w}_m\|^2 \sigma_{e,k}^2}{|\boldsymbol{w}_m^H \hat{\boldsymbol{h}}_{k,m}|^2 + \|\boldsymbol{w}_m\|^2 \sigma_{e,k}^2} + \|\boldsymbol{w}_m\|^2 \sigma_z^2 - K^2 \Gamma\Big)\),
whose first derivative for \(\{\tilde{b}_{k,m}\}\) equals zero when
\begin{equation} \label{32}
	\begin{aligned}
		\tilde{b}_{k,m} = \frac{\lambda_m |{\boldsymbol{w}_m}^H \hat{\boldsymbol{h}}_{k,m}|}{\lambda_m |{\boldsymbol{w}_m}^H \hat{\boldsymbol{h}}_{k,m}|^2 + \lambda_m \|\boldsymbol{w}_m\|^2 \sigma_{e,k}^2 + \mu_k} &, \\ \forall k \in \mathcal{K}, \mu_k > 0 &.
	\end{aligned}	
\end{equation}
Similar to Lemma 4 in Appendix B, there is always at least one WD \(k\) satisfying \(\mu_k > 0\) and it holds that \(\lambda_m > 0\).
Substituting \eqref{32} back, the constraint in (P4.2.\(m\)) w.r.t. \(\lambda_m\) becomes
\begin{equation} \label{41}
	\begin{aligned}
		\sum_{k \in \mathcal{K}} &\frac{(\lambda_m \|\boldsymbol{w}_m\|^2 \sigma_{e,k}^2 + \mu_k)^2 + \lambda_m^2 |{\boldsymbol{w}_m}^H \hat{\boldsymbol{h}}_{k,m}|^2 \|\boldsymbol{w}_m\|^2 \sigma_{e,k}^2}{(\lambda_m |{\boldsymbol{w}_m}^H \hat{\boldsymbol{h}}_{k,m}|^2 + \lambda_m \|\boldsymbol{w}_m\|^2 \sigma_{e,k}^2 + \mu_k)^2} \\
		 &\le K^2 \Gamma - \|\boldsymbol{w}_m\|^2 \sigma_z^2,
	\end{aligned} 
\end{equation}
whose L.H.S. monotonously decreases from \(K^2 \bar{\Gamma}_{m,2}\) to \(K^2 \bar{\Gamma}_{m,1}\) as \(\lambda_m\) increases from \(0\) to infinity, and the equality holds at the optimality as \(\lambda_m > 0\). Therefore, we can obtain the optimal dual variable \(\lambda_m^{**}\) by applying the bisection search based on the equality of \eqref{41}. Substituting \(\lambda_m^{**}\) back to \eqref{32}, we get the optimal solution to (P4.2.\(m\)), and thus (P4.1.\(m\)) as \(\{\tilde{b}_{k,m}^{**}\}\). This thus completes the proof.

\bibliographystyle{IEEEtran}
\bibliography{IEEEabrv,AirComp3}

\end{document}